\documentclass[a4paper,useAMS,usenatbib,usegraphicx]{mn2e}
\input{psfig.sty}
\usepackage{amsmath}
\usepackage{ulem}
\usepackage{color}
\newcommand{\kmsmpc}{\>{\rm km}\,{\rm s}^{-1}\,{\rm Mpc}^{-1}}

\newcommand{\mpch}{\>h^{-1}{\rm {Mpc}}}

\newcommand{\beq}{\begin{equation}}
\newcommand{\eeq}{\end{equation}}

\newcommand{\rmd}{{\rm d}}

\newcommand{\vect}[1]{{\bf #1}}
\newcommand{\bolds}[1]{{\boldsymbol #1}}
\newcommand{\dsbydt}[3]{{\frac{\partial^2 #3}{\partial #1\partial #2}}}
\newcommand{\dbyd}[2]{{\frac{\partial #1}{\partial #2}}}

\newcommand{\eV}{\>{\rm eV}}

\newcommand{\avg}[1]{\langle #1 \rangle}

\newcommand{\drm}{{\rm d}}

\def\gtsima{$\; \buildrel > \over \sim \;$}
\def\ltsima{$\; \buildrel < \over \sim \;$}
\def\prosima{$\; \buildrel \propto \over \sim \;$}
\def\gsim{\lower.7ex\hbox{\gtsima}}
\def\lsim{\lower.7ex\hbox{\ltsima}}
\def\simgt{\lower.7ex\hbox{\gtsima}}
\def\simlt{\lower.7ex\hbox{\ltsima}}
\def\simpr{\lower.7ex\hbox{\prosima}}

\def\ga{\gsim}

\def\gta{\ga}

\newcommand{\apj}{ApJ}

\newcommand{\apjs}{ApJS}
\newcommand{\aj}{AJ}
\newcommand{\mnras}{MNRAS}
\newcommand{\aap}{A\&A}

\newcommand{\prd}{Physical Review D}

\newdimen\hssize
\newdimen\hdsize
\def\rmb{{\rm b}}
\def\rmc{{\rm c}}
\def\rmd{{\rm d}}
\def\rme{{\rm e}}

\def\rmg{{\rm g}}
\def\rmh{{\rm h}}

\def\rmK{{\rm K}}

\def\rmm{{\rm m}}

\def\rmp{{\rm p}}

\def\rmr{{\rm r}}
\def\rms{{\rm s}}
\def\rmt{{\rm t}}

\def\rmx{{\rm x}}
\def\rmy{{\rm y}}

\def\rmK{{\rm K}}

\def\rmT{{\rm T}}

\def\rmV{{\rm V}}

\def\calH{{\cal H}}

\def\calL{{\cal L}}

\begin{document}
\setlength{\hbadness}{10000}
%%%%%%%%%%%%%%%%%%%%%%%%%%%%%%%%%%%%%%%%%%%%%%%%%%%%%%%%%%%%%%%%%%%%%%%%%%

\title[Cosmological Constraints from Clustering \& Lensing]
      {Cosmological Constraints from a Combination of Galaxy Clustering 
         \& Lensing -- II. Fisher Matrix Analysis}

\author[More et al.]
       {\parbox[t]{\textwidth}{
        Surhud More$^{1}$\thanks{E-mail: surhud@kicp.uchicago.edu}\thanks{KICP fellow}, 
        Frank C. van den Bosch$^{2}$, 
        Marcello Cacciato$^{3}$ \thanks{Minerva fellow},
        Anupreeta More$^{1}$, Houjun Mo$^4$, Xiaohu Yang$^5$}\\
           \vspace*{3pt} \\
   $^{1}$Kavli Institute for Cosmological Physics, University of Chicago,
	  933 East 56th Street, Chicago, IL 60637, USA \\
   $^{2}$Astronomy Department, Yale University , P.O. Box 208101, New
   Haven, CT 06520-8101, USA \\
   $^{3}$Racah Institute of Physics, The Hebrew University, Jerusalem
	  91904, Israel \\
    $^4$Department of Astronomy, University of Massachusetts,
            Amherst MA 01003-9305\\
    $^5$Key Laboratory for Research in Galaxies and Cosmology, Shanghai
        Astronomical Observatory, the Partner Group of MPA,\\
    $\;$Nandan Road 80, Shanghai 200030, China
       }

%%%%%%%%%%%%%%%%%%%%%%%%%%%%%%%%%%%%%%%%%%%%%%%%%%%%%%%%%%%%%%%%%%%%%%%%%%

\date{}

\maketitle

\label{firstpage}

%%%%%%%%%%%%%%%%%%%%%%%%%%%%%%%%%%%%%%%%%%%%%%%%%%%%%%%%%%%%%%%%%%%%%%%%%%

\begin{abstract}
We quantify the accuracy with which the cosmological parameters
characterizing the energy density of matter ($\Omega_\rmm$), the
amplitude of the power spectrum of matter fluctuations ($\sigma_8$),
the energy density of neutrinos ($\Omega_{\nu}$) and the dark energy
equation of state ($w_0$) can be constrained using data from large
galaxy redshift surveys. We advocate a joint analysis of the abundance
of galaxies, galaxy clustering, and the galaxy-galaxy weak lensing
signal in order to simultaneously constrain the halo occupation
statistics (i.e., galaxy bias) and the cosmological parameters of
interest.  We parameterize the halo occupation distribution of
galaxies in terms of the conditional luminosity function and use the
analytical framework of the halo model described in our companion
paper (van den Bosch et al.~2012), to predict the relevant
observables. By performing a Fisher matrix analysis, we show that a
joint analysis of these observables, even with the precision with
which they are currently measured from the Sloan Digital Sky Survey,
can be used to obtain tight constraints on the cosmological
parameters, fully marginalized over uncertainties in galaxy bias.  We
demonstrate that the cosmological constraints from such an analysis
are nearly uncorrelated with the halo occupation distribution
constraints, thus, minimizing the systematic impact of any
imperfections in modeling the halo occupation statistics on the
cosmological constraints.  In fact, we demonstrate that the
constraints from such an analysis are both complementary to and
competitive with existing constraints on these parameters from a
number of other techniques, such as cluster abundances, cosmic shear
and/or baryon acoustic oscillations, thus paving the way to test the
concordance cosmological model.
\end{abstract}
%We present the Fisher information content in each of
%these observables and 

%%%%%%%%%%%%%%%%%%%%%%%%%%%%%%%%%%%%%%%%%%%%%%%%%%%%%%%%%%%%%%%%%%%%%%%%%%


%%%%%%%%%%%%%%%%%%%%%%%%%%%%%%%%%%%%%%%%%%%%%%%%%%%%%%%%%%%%%%%%%%%%%%%%%%
\section{Introduction} 
\label{sec:intro}

Of the various cosmological models in the literature, the $\Lambda$CDM
model has withstood a large variety of observational tests which have
grown more and more stringent over time. According to this model, at
present times, dark energy and dark matter form the dominant component
of the energy density budget of the Universe. Ordinary matter forms a
sub-dominant component and is primarily present in the form of gas in
the intergalactic medium and around galaxies. Obtaining precise
constraints on the energy densities of the various components of the
Universe (quantified by the energy density parameters for matter
[$\Omega_\rmm$], dark energy [$\Omega_{\Lambda}$], baryonic matter
[$\Omega_\rmb$] and neutrinos [$\Omega_\nu$]), is of great
importance, in order to understand the expansion history and future
fate of the Universe.

Any successful cosmological model also requires to explain the
formation of structure in the Universe. The $\Lambda$CDM model assumes
the presence of nearly scale invariant, tiny initial fluctuations in
the matter density, presumably generated during inflation. Structure grows
hierarchically in this model with small scales collapsing first
followed by larger and larger scales as the Universe grows old. The
precise description of the inhomogeneous Universe requires the
knowledge of the index of the power spectrum of initial fluctuations,
$n_\rms$ and its amplitude (quantified in terms of the parameter
$\sigma_8$).

Observations of the fluctuations in the cosmic microwave background on
both large \citep[e.g.,][]{Spergel2007,Komatsu2011} and small scales
\citep[e.g.,][]{Lueker2010,Dunkley2011}, of the dimming of distant
supernovae as a function of their redshift \citep[e.g.,][]{Riess1998,
  Perlmutter1999,Kowalski2008, Kessler2009, Guy2010}, of the large
scale structure traced by galaxies
\citep[e.g.,][]{Tegmark2004,Swanson2010}, of the abundance of massive
clusters and its evolution \citep[e.g.,][]{Vikhlinin2009,Mantz2010,Rozo2010,Benson2011,Sehgal2011}, of the
angular extent of the baryonic acoustic oscillation scale as a
function of redshift
\citep[e.g.,][]{Eisenstein2005,Percival2007,Percival2010,Blake2011,Anderson2012}, and
of the weak distortion of galaxies due to intervening structure
\citep[e.g.,][]{Massey2007,Schrabback2010,Lin2011,Huff2011} have all contributed to
enable precise constraints on the parameters that describe the
$\Lambda$CDM model. It is remarkable that a single model with
a small number of parameters is able to self consistently explain all
these observations.

Large spectroscopic galaxy redshift surveys such as the 2-degree field
galaxy redshift survey \citep{Colless2001} and the Sloan digital sky
survey \citep[hereafter SDSS,][]{York2000,Abazajian2009} have played an
important role of mapping out the three-dimensional structure of
galaxies in the Universe in exquisite detail. This has allowed precise
measurements of the abundance of galaxies and their clustering on
non-linear scales as a function of galaxy properties such as
luminosity \citep{Wang2008,Zehavi2011,Blanton2003b,Norberg2001},
stellar mass \citep{Li2007,Baldry2008,Cole2001}, or colour 
\citep{Norberg2002,Wang2008,Swanson2008,Zehavi2011}. The halo model
has traditionally been used to obtain the halo occupation distribution
of galaxies given a cosmological model based on these observations
\citep[e.g.,][]{Jing1998,Berlind2002, Yang2003, Zheng2005, vdb2007,
  Zheng2007, Tinker2005, Zehavi2011,Coupon2011,Leauthaud2012}. A few
studies however have also turned the question around to judge if these
observations can be used to constrain the cosmological parameters
themselves \citep{vdb2003, vdb2007, Tinker2005, Yoo2006, Cacciato2009,
  Tinker2012}.

Galaxies reside in dark matter haloes. The cosmological parameters
determine the abundance and clustering of dark matter haloes of a
given mass. Therefore, the observations of the abundance and
clustering of galaxies is sensitive to various cosmological parameters
and can be used to constrain these parameters. However, this requires
the knowledge of an accurate mapping between galaxies and their dark
matter haloes. In Cacciato et al.  (2009), we highlighted this
problem, by demonstrating that two different cosmological models,
(differing primarily in $\Omega_\rmm$ and $\sigma_8$) were able to
simultaneously fit the galaxy abundance and the galaxy-galaxy
clustering data by adjusting the halo occupation distribution of
galaxies. However, we also showed that these models make vastly
different predictions for the amount of baryons that should occupy
dark matter haloes, quantified by the mass-to-light ratio \citep[see
also][]{vdb2003, vdb2007, Tinker2005, Yoo2006}. Galaxy-galaxy lensing,
which measures the projected galaxy-matter correlation function
is an excellent probe of the mass-to-light ratios. Therefore, it was
suggested that a joint analysis of the abundance, clustering and
lensing of galaxies can be used to obtain constraints on cosmological
parameters such as $\Omega_\rmm$ and $\sigma_8$.

This paper is the second in a series of three papers in which we
develop this idea further. In van den Bosch et al. (2012, hereafter
Paper I) we describe our model for the halo occupation distribution of
galaxies and present an analytical framework for the computation of
the abundance of galaxies, their clustering and the galaxy-matter
cross correlation. We make extensive use of mock catalogs to validate
our analytical prescription and show that our analytical model is
accurate to better than 10 (in most cases 5) percent in reproducing
the 3-dimensional galaxy-galaxy correlation and galaxy-matter
correlation. These correlations can then be projected to predict the
galaxy clustering and the galaxy-galaxy lensing signal measured from
SDSS data. In this paper, we perform a Fisher matrix analysis to
forecast the accuracy with which various cosmological parameters can
be constrained with current data. The analysis allows us to understand
the strength of each of our data-sets as well as the
various degeneracies between our model parameters. In Cacciato et al.
(2012, hereafter Paper III), we apply our method to data from the
SDSS, simultaneously constraining galaxy bias and
cosmology\footnote{A preliminary version of the main results from
this paper and Paper III were published in a conference
proceedings by \citet{More2012clf}.}.

This paper is organized as follows. In section \ref{sec:data}, we
briefly describe the data and our analytical model. In section
\ref{sec:Fisher}, we describe the Fisher information matrix
corresponding to the observables. In section \ref{sec:constraints}, we
present the constraints on our model parameters that are statistically
achievable given the accuracy of the data. Finally, we summarize our
results in Section \ref{sec:summary} and provide a future outlook.

\section{Data and the analytic model} 
\label{sec:data}

Our analysis is based on galaxy observations carried out as part of
the spectroscopic component of the SDSS. In particular, we focus on
the accurate measurements of the galaxy luminosity function
\citep{Blanton2003a}, the clustering of galaxies in six different
luminosity bins \citep{Zehavi2011} on scales ranging from $0.17\mpch$
to $42.3\mpch$, and the galaxy-galaxy lensing signal around galaxies
in different luminosity bins \citep{Mandelbaum2006} on scales ranging
from $0.045\mpch$ to $1.81\mpch$. Our aim is to perform a Fisher
matrix analysis to understand the extent to which these datasets
constrain the cosmological parameters that we are interested in and
the various degeneracies that enter our analysis. For this purpose, we
choose a set of fiducial parameters that describe the cosmology (see
Table~\ref{tab:tab1}) and the halo occupation distribution of the
galaxies (see Table~\ref{tab:tab2}). We use the analytical model (see
Sections~\ref{sec:CLF}-\ref{sec:gglensing}) to predict galaxy
abundances, galaxy clustering (in projection), and the galaxy-galaxy
lensing signal (always using the same luminosity bins and radii as in
the SDSS data). We assume that these observables are measured with a
percentage accuracy that is listed in Table~\ref{tab:tab3} and chosen
such that the measurement quality is fairly similar to that of actual
SDSS measurements.

We perform three different analyses, that correspond to the
$\Lambda$CDM model and its variants. For the first analysis, we assume
a flat standard $\Lambda$CDM cosmology with a negligible amount of
energy density in neutrinos. We primarily focus on the matter
density parameter, $\Omega_\rmm$ and the parameter that characterizes
the amplitude of matter fluctuations, $\sigma_8$. We use priors on the
secondary cosmological parameters, the baryon energy density parameter
($\Omega_\rmb$), the Hubble parameter ($h=H_0/100\kmsmpc$) and the
power law index of the initial power spectrum of density fluctuations
($n_\rms$) from the analysis of the 7 year data from WMAP
\citep{Komatsu2011}.  This is the fiducial model and set of priors
which will also be used in Paper III. For the second analysis, we allow for
the presence of a non-zero neutrino mass
which gives rise to a non-zero neutrino density parameter,
$\Omega_{\nu}$.  We assume uninformative prior information about $\Omega_{\nu}$
(except that it is positive-definite) and also continue to use
uninformative
priors on $\Omega_\rmm$ or $\sigma_8$. For the third analysis, we
restore our assumption of a negligible amount of neutrino energy
density, but this time add the parameter $w_0$ that describes the dark
energy equation of state (EoS), such that
\begin{equation} w_0=P_{\Lambda}/(c^2\,\rho_{\Lambda})\, .
\end{equation}
Here $P_{\Lambda}$ describes the pressure exerted by dark energy and
$\rho_{\Lambda}$ denotes its energy density.  

\begin{table} 
\caption{Cosmological model parameters} 
\begin{center} 
\begin{tabular}{clll} 
\hline
\hline 
Cosmological & Model A & Model B & Model C \\ 
parameters   &         &         &          \\
\hline
$\Omega_\rmm$       & ~0.266        &  ~0.266       & ~0.266        \\
$\Omega_\rmK$       & ~0.0$^{\dagger}$   &  ~0.0$^{\dagger}$  &
~0.0$^{\dagger}$   \\
$\Omega_\rmb h^2$       & ~0.02258$^{*}$  &  ~0.02258$^{*}$ & ~0.02258$^{*}$  \\
$\Omega{\nu}$    & ~0.0$^{\dagger}$   & ~0.004         & ~0.0$^{\dagger}$   \\
$\sigma_8$       & ~0.801         &  ~0.801        & ~0.801         \\
$n_\rms$            & ~0.963$^{*}$  &  ~0.963$^{*}$ & ~0.963$^{*}$  \\
$h$              & ~0.71$^{*}$    &  ~0.71$^{*}$   & ~0.71$^{*}$  \\
$w_0$            & -1.0$^{\dagger}$   &  -1.0$^{\dagger}$  & -1.0         \\
\hline
\hline 
\end{tabular} 
\end{center}
\begin{minipage}
{\hsize} 
Cosmological model parameters used to generate artificial datasets.
We use a prior given by the covariance matrix of the WMAP chain for
the parameters marked with an asterisk. The parameters marked with a
dagger are kept fixed when analysing the model for the fiducial case.
We discuss variations on some of the assumed priors in the text.
\end{minipage} 
\label{tab:tab1} 
\end{table}

\begin{table} 
\caption{CLF and nuisance model parameters} 
\begin{center} 
\begin{tabular}{clllll} 
\hline
\hline 
Central CLF   & x$L_0$ & x$M_1$ & $\gamma_1$ & $\gamma_2$ & $\sigma_\rmc$ \\ 
\hline
              & 9.93   & 11.19  & 3.5        & 0.25       & 0.156       \\
\hline
\hline
Satellite CLF & $b_0$ & $b_1$ & $b_2$ & $\alpha_\rms$ &    \\
\hline
              & -1.08  & 1.46   & -0.20      & -1.15       & \\
\hline
\hline 
Nuisance & ${\eta}$ & $\psi$  &  &  &    \\
\hline
              & 1.0  & 0.903  &  &  & \\
\hline
\hline 
\end{tabular} 
\end{center}
\begin{minipage}
{\hsize} 
Model parameters used to describe the central and satellite CLFs.
Bottom row shows the values of the nuisance parameters in our model
that we assumed for the Fisher analysis.
\end{minipage} 
\label{tab:tab2} 
\end{table}

\begin{table} 
\caption{Luminosity bins of the SDSS clustering data} 
\begin{center} 
\begin{tabular}{cccc} 
\hline
\hline 
$^{0.1}M_\rmr - 5 \log h$ & $\langle z \rangle$ & $w_\rmp$ accuracy &
$\Delta\Sigma$ accuracy \\ 
(1)                    & (2)                 &  (3)        & (4) \\
\hline
$(-19.0,-18.0]$ & 0.047 & 30\% & 50\% \\ 
$(-20.0,-19.0]$ & 0.071 & 30\% & 20\% \\ 
$(-21.0,-20.0]$ & 0.10  & ~5\% & 15\% \\ 
$(-21.5,-21.0]$ & 0.14  & ~5\% & 15\% \\ 
$(-22.0,-21.5]$ & 0.17  & ~5\% & 15\% \\
$(-22.5,-22.0]$ & 0.20  & 40\% & 15\% \\ 
\hline
\hline 
\end{tabular} 
\end{center}
\begin{minipage}
{\hsize} 
Luminosity  bins. Column (1) indicates the magnitude range of each
luminosity bin (all magnitudes are K$+$E corrected to $z=0.1$). Column
(2) indicates the mean redshift of the lens galaxies  in each
luminosity bin. Column (3) indicates the accuracy of the galaxy
clustering data and column (4) indicates the accuracy of the
galaxy-galaxy lensing data. The luminosity function data have a
fractional accuracy of $\sim 4\%$, and this accuracy degrades to about
$\sim 10-15\%$ at the bright and the faint ends respectively.
\end{minipage} 
\label{tab:tab3} 
\end{table}

\subsection{The Conditional Luminosity Function}
\label{sec:CLF}

We use the conditional luminosity function (CLF) model described in
Paper I to specify our halo occupation distribution. Readers familiar
with our model from Paper I may want to proceed directly to Section
\ref{sec:Fisher}.

The CLF, $\Phi(L|M)\,\rmd L$, is defined to be the average number of
galaxies with a luminosity $L\pm \rmd L/2$ that reside in a halo of
mass $M$ \citep{Yang2003}. The CLF consists of a contribution from 
central galaxies, $\Phi_\rmc(L|M)$, and from satellite galaxies,
$\Phi_\rms(L|M)$.  The distribution $\Phi_\rmc(L|M)$ is described by a
lognormal distribution with a scatter, $\sigma_\rmc$, that is
independent of halo mass, consistent with the findings from studies of
satellite kinematics \citep{More2009b, More2009a, More2011a} and
galaxy group catalogues \citep{Yang2009}. The dependence of the
logarithmic mean luminosity, $\log \tilde{L}_\rmc$, on halo mass is
given by
\begin{equation}
\log \tilde{L}_\rmc(M)=\log \left[ L_0
\frac{(M/M_1)^{\gamma_1}}{\left[1+(M/M_1)\right]^{\gamma_1-\gamma_2}}
\right]\,.
\end{equation}
Four parameters are required to describe this dependence; two
normalization parameters, $L_0$ and $M_1$ \footnote{The x in front of
  the $L_0$ and $M_1$ in Table~\ref{tab:tab2} indicates that the
  parameters we use in practice are 10-based logarithm of $L_0$ and
  $M_1$, respectively.} and two parameters $\gamma_1$ and $\gamma_2$
that describe the slope of the $\tilde{L}_\rmc(M)$ relation at the low
mass end and the high mass end, respectively.

The satellite CLF, $\Phi_\rms(L|M)$ is assumed to be a
Schechter-like function, 
\begin{equation}
\Phi_\rms(L|M) \drm L=\Phi_\rms^*\left(\frac{L}{L_*}\right)^{\alpha_\rms}\,
\exp\left[-\left( \frac{L}{L_*} \right)^2 \right] \,\frac{\drm L}{L_*}.
\end{equation}
Here $L_*(M)$ determines the knee of the satellite CLF and is assumed
to be a factor $f_\rms$ times fainter than $\tilde{L}_\rmc(M)$.
Motivated by results from the SDSS group catalog of \citet{Yang2008a},
we set $f_\rms = 0.562$ and assume that the faint-end slope of the
satellite CLF is independent of halo mass.  The logarithm of the
normalization, $\Phi_\rms^*$ is assumed to have a quadratic dependence
on $\log M$ described by three free parameters, $b_0$, $b_1$ and
$b_2$;
\begin{equation}
\log \Phi_\rms^*=b_0+b_1\,(\log M-12)+b_2\,(\log M-12)^2\,.
\end{equation}
Note that this functional form does not have a physical motivation; it
merely provides an adequate description of the results obtained by
\citet{Yang2008a} from the SDSS galaxy group catalog. In addition to
specifying the luminosity dependence of the halo occupation
distribution, we also need to specify the spatial distribution of
galaxies in dark matter haloes. Throughout, we assume that central
galaxies reside at the center of their haloes and that the satellite
galaxies follow the matter distribution without any spatial bias.
The matter density distribution is given by the NFW profile
\citep{Navarro1997} with a concentration that depends upon mass
according to the calibration 
presented in \citet{Maccio2007}. As described in Paper I, we allow a
$10$ percent uncertainty in the calibration of the normalization of
the concentration-mass relation to account for our neglect of the
baryonic effects on the matter distribution of halos and the scatter in
the concentration-mass relation. We express this uncertainty as a
multiplicative parameter ${\eta}$ to the fiducial concentration-mass
relation and marginalize over this nuisance parameter. We use the
parameters listed in Table~\ref{tab:tab2} for all the three
cosmological models.

In all our models the abundance and clustering of dark matter haloes
is set by the cosmological parameters, while the halo occupation
distribution, as parameterized by the CLF, describes how this
abundance and clustering of haloes translates into the abundance and
clustering of galaxies as well as their cross correlation with
matter. In what follows, we briefly describe the expressions that can
be used to compute the model predictions for the data described above.

\subsection{Galaxy Luminosity Function}
\label{sec:phi}

Given the cosmological parameters and the parameters of the CLF, the
luminosity function of galaxies, $\Phi(L,z)$, simply follows from
multiplying the average number of galaxies in a halo of given mass
with the number densities of haloes of that mass, $n(M,z)\rmd M$, and by
integrating this product over all halo masses,
\begin{eqnarray}
\Phi(L,z) &=& \int_0^\infty \Phi(L|M)\,n(M,z)\,\drm M \,, \\
        &=& \int_0^\infty \left[  \Phi_\rmc(L|M) + \Phi_\rms(L|M)
            \right] \,n(M,z)\,\drm M \,.
\end{eqnarray}
It is clear from the above equation, that all of the cosmological
information in the galaxy luminosity function is due to its dependence
on the halo mass function. 

The fraction of galaxies of a particular luminosity that are
satellites is important to understand the relative strength of the
luminosity function to constrain the central CLF parameters compared
to that of the satellite CLF parameters.  The satellite fraction can
be obtained in the CLF formalism using
\begin{equation}
f_{\rm sat}(L,z) = \frac{1}{\Phi(L,z)} \int_{0}^{\infty} \Phi_\rms(L|M)
\,n(M,z)\,\drm M \,.
\end{equation}
The satellite fraction is also very crucial in determining the shape
of the galaxy-galaxy and galaxy-matter clustering signals
\citep[e.g.,][]{Seljak2005,Mandelbaum2006,Zehavi2011}. Typically,
the central galaxies dominate the galaxy population at all
luminosities under consideration. For the fiducial parameters that we
adopt for our analysis, the satellite fraction is very low at the
bright end, but increases to $30-40\%$ at the faint end, in good
agreement with observational constraints
\citep[e.g.,][]{Mandelbaum2006,vdb2007,Tinker2007}.

For the purpose of computing the galaxy-galaxy clustering and the
galaxy-galaxy lensing signal, we will be concerned with galaxies in a 
specific luminosity interval $[L_1,L_2]$. The average number density
of such galaxies follows from the CLF according to
\begin{equation}\label{avgngal}
\bar{n}_\rmg(z) = \int \langle N_\rmg|M \rangle \, n(M,z) \, \rmd M\,,
\end{equation}
where
\begin{equation}\label{avgnm}
\langle N_\rmg|M\rangle = \int_{L_1}^{L_2} \Phi(L|M) \rmd L\,,
\end{equation}
is the average number of galaxies with $L_1 < L < L_2$ that reside in
a halo of mass $M$.

\subsection{Galaxy Clustering}
\label{sec:ggclustering}

The clustering of galaxies in the SDSS data is measured using a volume
limited sample of galaxies and is expressed in terms of the
galaxy-galaxy correlation function. The correlation function expresses
the excess probability over random to find a pair of galaxies
separated by a given distance. The galaxy-galaxy correlation
function consists of two different terms based upon the kind of galaxy
pairs under consideration. The correlation function due to pairs of
galaxies that reside within the same dark matter halo is called the
one-halo term. On the other hand, the correlation function due to
pairs of galaxies that reside in separate dark matter haloes is called
the two-halo term.

The one-halo correlation can be further subdivided into the
central-satellite term and satellite-satellite term based on the kind
of galaxies that constitute the pair. Similarly the two-halo
correlation can be subdivided into the central-central term, the
central-satellite term and the satellite-satellite term.  For
computational simplicity, each of these correlation function terms are
computed in Fourier space. The power spectrum and the correlation
function form a Fourier transform pair. 

The galaxy-galaxy power spectrum, $P_{\rm gg}(k,z)$ can be expressed as
the sum of the following terms,
\begin{eqnarray}
P_{\rm gg}(k,z) &=& 2\,P^{\rm 1h}_{\rm cs}(k,z) + P^{\rm 1h}_{\rm
ss}(k,z) \nonumber \\ && + P^{\rm 2h}_{\rm cc}(k,z) + 2\,P^{\rm
2h}_{\rm cs}(k,z) + P^{\rm 2h}_{\rm ss}(k,z)\,.
\end{eqnarray}
As shown in Paper I, these terms can be written in a compact form as
\begin{equation}\label{P1h}
P^{\rm 1h}_{\rm xy}(k,z) = \int \calH_\rmx(k,M,z) \, \calH_\rmy(k,M,z) \, 
n(M,z) \, \rmd M,
\end{equation}
\begin{eqnarray}\label{P2h}
\lefteqn{P^{\rm 2h}_{\rmx\rmy}(k,z) =
\int \rmd M_1 \, \calH_\rmx(k,M_1,z) \, n(M_1,z) } \nonumber \\
& & \times \int \rmd M_2 \, \calH_\rmy(k,M_2,z) \, n(M_2,z) \,
Q(k|M_1,M_2,z)\,,
\end{eqnarray}
where `x' and `y' are either `c' (for central) or `s' (for satellite),
$Q(k|M_1,M_2,z)$ describes the power-spectrum of haloes of masses $M_1$
and $M_2$, and we have defined
\begin{equation}\label{calHc}
\calH_\rmc(k,M,z) = \calH_\rmc(M,z) = 
{\langle N_\rmc|M \rangle \over \bar{n}_{\rmg}(z)} \,,
\end{equation}
and
\begin{equation}\label{calHs}
\calH_\rms(k,M,z) = {\langle N_\rms|M \rangle \over \bar{n}_{\rmg}(z)} \,  
\tilde{u}_\rms(k|M,z)\,.
\end{equation}
Here $\langle N_\rmc|M \rangle$ and $\langle N_\rms|M \rangle$ are the
average number of central and satellite galaxies in a halo of mass
$M$, which follow from Eq.~(\ref{avgnm}) upon replacing $\Phi(L|M)$ by
$\Phi_\rmc(L|M)$ and $\Phi_\rms(L|M)$, respectively.  Furthermore,
$\tilde{u}_\rms(k|M)$ is the Fourier transform of the normalized
number density distribution of satellite galaxies that reside in a
halo of mass $M$.

The power spectrum of haloes $Q(k|M_1,M_2,z)$ is the most uncertain
component of the halo model, as it needs to account for the large
scale bias of haloes and the corresponding scale dependence as well as
halo exclusion (i.e., the fact that haloes are spatially mutually
exclusive). We do not describe the details of our treatment here but
simply refer the interested reader to Paper I.  We emphasize, though,
that our analytical model for $Q(k|M_1,M_2,z)$ has been tested and
calibrated using high-resolution numerical simulations.  Nevertheless,
to account for uncertainties, we promote one of the calibration
parameters, $\psi$, which is used to characterize the scale dependence
of the halo bias, to be a part of our parameter set. Throughout we
adopt a $15$ percent uncertainty on $\psi$.

The corresponding galaxy-galaxy correlation function,
$\xi_{\rm gg}(r,z)$, can be obtained via a Fourier transform of the
galaxy-galaxy power spectrum,
\begin{equation}
\xi_{\rm gg}(r,z)
 = 
{1 \over 2 \pi^2} \int_0^{\infty} P_{\rm gg}(k,z) {\sin kr \over kr}
\, k^2 \rmd k\,.
\end{equation}
However, $\xi_{\rm gg}(r,z)$ cannot be measured directly from
observations. This is a direct consequence of our inability to infer
the line-of-sight separation of galaxies from redshift surveys due to
the peculiar motions of galaxies. Instead, the correlation function of
galaxies is measured separately as a function of the separation of
galaxies along the line-of-sight and in the plane of the sky to obtain
the redshift space correlation function $\xi_{\rm
  gg}^z(r_\rmp,r_\pi,z)$. This function is then integrated along the
line-of-sight to obtain the projected correlation function
$w_\rmp(r_\rmp,z)$,
\begin{equation}
w_\rmp(r_\rmp,z) = 2 \int_{0}^{r_{\rm max}} \xi^z_{\rm
gg}(r_\rmp,r_\pi,z)
\rmd r_\pi\,.
\end{equation}
Here $r_{\rm max}$ is the maximum line-of-sight distance to which
the redshift space correlation function is integrated in order to
obtain $w_\rmp(r_\rmp,z)$. Only for $r_{\rm max}=\infty$, the projected
correlation function is entirely independent of redshift space
distortions, and can therefore be expressed in terms of the real space
correlation function according to
\begin{equation}
w_\rmp(r_\rmp,z)=\int_{r_\rmp}^{\infty} \xi_{\rm
gg}(r,z) \frac{2\,r\,\rmd r}{\sqrt{r^2-r_\rmp^2}}
\end{equation}
However, since real data sets are always limited in extent, in
practice the projected correlation function $w_\rmp(r_\rmp,z)$ is always
obtained by integrating $\xi_{\rm gg}^z(r_\rmp,r_{\pi})$ out to some
finite $r_{\rm max}$ rather than to infinity. For example,
\citet{Zehavi2011}, whose data we use in Paper~III, adopt $r_{\rm max}
= 40\,\mpch$ or $60\,\mpch$, depending on the luminosity
sample used. As discussed in paper~I, such values of $r_{\rm max}$ are
sufficient to get rid of the small scale redshift space distortions
(commonly called the Finger-of-god effect) but the data suffer from
residual redshift space distortions on large scales that, unless
corrected for, can easily result in systematic errors of 10 percent or
larger \citep[see also][]{Norberg2009,Baldauf2010,More2011b}. In Paper
I, we have shown that a slightly modified version of the model
presented by \citet{Kaiser1987} can be used to correct for these
residual redshift space distortions. Tests using mock galaxy redshift
surveys show that this method is accurate at the few percent level.
We refer the interested reader to Paper~I for further details. 

\subsection{Galaxy- Galaxy Lensing}
\label{sec:gglensing}

The clustering of matter around galaxies (i.e., the galaxy-matter
cross correlation) can be probed using galaxy-galaxy lensing; the
small distortions of the shapes of background galaxies (the sources)
due to weak gravitational lensing by the matter surrounding
foreground galaxies (the lenses).  Since these shape distortions are
extremely weak, background galaxies have non-zero ellipticities, and
one typically can only identify a few background galaxies per lens
galaxy, a large number of lens galaxies need to be stacked in order to
detect a signal. The average tangential ellipticity,
$\avg{\epsilon_\rmt}$, around a stack of foreground galaxies is equal
to the tangential shear, $\gamma_\rmt$, which is related to the
projected surface density around the lens galaxies according to
\begin{equation}
\avg{\epsilon_\rmt}(R,z)=\gamma_\rmt(R,z)= \frac{\Delta \Sigma(R,z)}{\Sigma_{\rm
crit}}=\frac{\bar\Sigma(R,z)-\Sigma(R,z)}{\Sigma_{\rm crit}} \,.
\label{eq:dsigma}
\end{equation}
Here, $\bar\Sigma(R,z)$ is the projected matter density averaged within
a circular aperture of radius $R$ centered around the foreground lens
galaxy at redshift $z$ and $\Sigma(R,z)$ is the projected surface density
at a distance $R$ from the lens galaxy.  The critical surface density
$\Sigma_{\rm crit}$ is a geometric factor that includes the angular
diameter distances from the observer to the source, the observer to
the lens and the lens to the source. The projected matter density
around the foreground lens galaxy can be calculated from the
galaxy-matter cross correlation, $\xi_{\rm gm}(r,z)$, using
\begin{equation}
\Sigma(R,z) = \int_{R}^{\infty} \bar{\rho}\,[1+\xi_{\rm gm}(r,z)]
\frac{2\,r\,\drm r}{\sqrt{r^2-R^2}}\,.
\label{eq:sigproj}
\end{equation}
For the purpose of calculating $\Delta \Sigma$, one can safely replace
$[1+\xi_{\rm gm}]$ with $\xi_{\rm gm}$ in the above equation.

The galaxy-matter cross correlation function can be modelled using the
conditional luminosity function in a manner that is very similar to
modelling the galaxy-galaxy correlation function: the clustering of
matter around galaxies can be divided into one-halo and two-halo
terms, each of which can be further subdivided into central and
satellite terms. We calculate each of these terms in Fourier space.
The total galaxy-matter power spectrum is given by the following sum
\begin{equation}\label{Pgm}
P_{\rm gm}(k,z) = P^{\rm 1h}_{\rm cm}(k,z) + P^{\rm 1h}_{\rm sm}(k,z) 
+ P^{\rm 2h}_{\rm cm}(k,z) + P^{\rm 2h}_{\rm sm}(k,z)\,.
\end{equation} 
The above terms can be calculated using Eqs.~(\ref{P1h})-(\ref{P2h}),
where `x' is `m' (for matter) and `y' is either `c' (for central) or
`s' (for satellite). For the matter component, we define
\begin{equation}\label{calHm}
\calH_\rmm(k,M,z) = {M \over \bar{\rho}_{\rmm}(z)} \,
\tilde{u}_\rmh(k|M,z)\,,
\end{equation}
where $\tilde{u}_\rmh(k|M,z)$ is the Fourier transform of the normalized
density distribution of matter within a halo of mass $M$ and
$\bar{\rho}_{\rmm}(z)$ denotes the average comoving density of the
Universe at redshift $z$. The
expressions for ${\cal H}_\rmc(k,M,z)$ and ${\cal H}_\rms(k,M,z)$ are given
by Eqs.~(\ref{calHc}) and (\ref{calHs}), respectively. The
galaxy-matter correlation function can be obtained by a Fourier
transform of the power spectrum given by Eq.~(\ref{Pgm}). The
galaxy-galaxy lensing signal, in turn, can be calculated using
Eqs.~(\ref{eq:dsigma}) and (\ref{eq:sigproj}).

The analytical model sketched above can be used to predict the
luminosity function of galaxies, their clustering strength and the
galaxy-galaxy lensing signal around them, given a set of cosmological
parameters and CLF parameters. These parameters can be constrained by
comparing the model prediction to observational data. In what follows,
we describe the sensitivity of these observations to various
parameters of our model, and forecast the accuracy with which existing
data from the SDSS can constrain our model parameters, in particular
those which describe the cosmology.

\section{Fisher information matrix}
\label{sec:Fisher}

All the observables in the datasets described above can be packed in a
single data vector denoted by $\vect{x}$. Our model consists of
parameters that describe how galaxies populate dark matter haloes, in
addition to the cosmological parameters that describe the statistics
of these dark matter haloes. We use the vector $\bolds{\theta}$ to
represent the set of our model parameters. Under the assumption that
the probability distribution of the data is a Gaussian, the
likelihood, $\calL$, of the data is given by
\begin{equation}
\ln {\cal L} = -\frac{1}{2}\ln |\vect{C}|
-\frac{1}{2}\,(\vect{x}-\bolds{\mu})^{\rm
T}\,\vect{C}^{-1}\,(\vect{x}-\bolds{\mu})\,.
\end{equation}
Here $\bolds{\mu}$ denotes the model predictions at the true parameter
value $\bolds{\theta}$ and $\vect{C}$ denotes the covariance matrix of
the observables. The Fisher information matrix $\vect{F}$ is given by
\begin{equation}
F_{ij} = -\left\langle\left.\dsbydt{\theta_i}{\theta_j}{\ln {\cal L}}
\right|_{\bolds{\theta}}\right\rangle\,,
\end{equation}
with the derivatives evaluated at $\bolds{\theta}$ and the angular
brackets denoting an ensemble average over all possible data
realizations.  Carrying out the differentiation of $\ln\calL$, we
obtain
\begin{eqnarray}
F_{ij}\!\!\!\!
&=&\!\!-\frac{1}{2}\left\langle \frac{\partial}{\partial \theta_j}\left[
  \frac{\partial \bolds{\mu}}{\partial
  \theta_i}^T\vect{C}^{-1}(\vect{x}-\bolds{\mu}) 
  +(\vect{x}-\bolds{\mu})^T\vect{C}^{-1}\frac{\partial \bolds{\mu}}{\partial \theta_i} 
  \right] \right\rangle \nonumber \\ \\
&=&\!\!\!\!-\frac{1}{2}\left[
  \frac{\partial^2 \bolds{\mu}}{\partial \theta_j
  \theta_i}^T\vect{C}^{-1}(\avg{\vect{x}}-\bolds{\mu}) 
  +(\avg{\vect{x}}-\bolds{\mu})^T\vect{C}^{-1}\frac{\partial^2
  \bolds{\mu}}{\partial \theta_i \partial \theta_j} \right] \nonumber \\
  &&+\frac{1}{2}\left[\frac{\partial \bolds{\mu}}{\partial
  \theta_i}^T\vect{C}^{-1}\frac{\partial \bolds{\mu}}{\partial
  \theta_j}
  +\frac{\partial \bolds{\mu}}{\partial
  \theta_j}\vect{C}^{-1}\frac{\partial \bolds{\mu}^T}{\partial
  \theta_i}\right]\,.
\end{eqnarray}
The terms in the first square bracket are zero because
$\avg{\vect{x}}=\bolds{\mu}(\bolds{\theta})$ and the terms in the
second bracket are equal because the covariance matrix is symmetric.
Therefore, we obtain
\begin{equation}
F_{ij}
=\frac{\partial \bolds{\mu}}{\partial
  \theta_i}^\rmT\vect{C}^{-1}\frac{\partial \bolds{\mu}}{\partial
    \theta_j} \,.
%\label{eq:fish}
\end{equation}
The inverse of the Fisher matrix represents the covariance of the
posterior probability distribution of the model parameters attainable
given the errorbars on the observables in the high signal-to-noise
ratio (SNR) limit (see \citealt{Vallisneri2008} for a more detailed
discussion). In particular, the Cramer-Rao inequality states that an
unbiased estimate of the parameter $\theta_i$ given the data can be
obtained with an errorbar $\sigma_i$ such that
\begin{equation}
\sigma_i \ge [(F^{-1})_{ii}]\,,
\end{equation}
when marginalized over the rest of the parameter set and where the
equality is satisfied in the high SNR limit.

For simplicity, we work with dimensionless model parameters,
$\lambda_i$, defined by
\begin{equation}
\theta_i=\lambda_i\,\tilde{\theta}_i\,,
\label{eq:dimpar}
\end{equation}
where $\tilde{\theta_i}$ denotes the maximum likelihood value for the
parameter $\theta_i$. The Fisher information matrix calculated using
these new variables is denoted by $\tilde{F}_{ij}$ and is related to
$F_{ij}$ according to
\begin{equation}
    \tilde{F}_{ij}=  \dbyd{\bmu}{\lambda_i}^\rmT \vect{C}^{-1}
\dbyd{\bmu}{\lambda_j} = \tilde{\theta}_i \tilde{\theta}_j
F_{ij} \,.
\label{eq:fish2}
\end{equation}
With this change of variables, the Fisher information matrix,
$\tilde{\vect{F}}$ is also dimensionless and the accuracy with which
$\lambda_i$ can be constrained denotes the {\it fractional} accuracy
with which the parameter $\theta_i$ can be constrained given the data.
In case of a diagonal covariance matrix, the dimensionless Fisher
matrix is given by
\begin{equation}
\tilde{F}_{ij}= \sum_k \frac{1}{C_{kk}} \left( \dbyd{\mu_k}{\lambda_i}
\dbyd{\mu_k}{\lambda_j} \right) \,,
\label{eq:fish}
\end{equation}
where the summation is over all data points. 

\begin{figure*}
\centerline{\psfig{figure=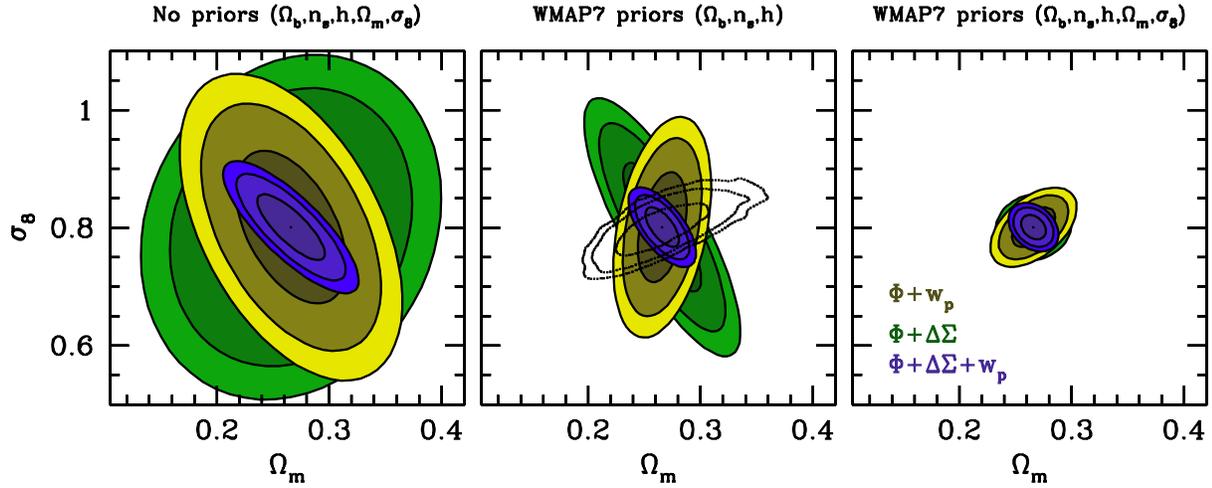,width=0.9\hdsize}}
\caption{ Cosmological constraints in the $\Omega_\rmm-\sigma_8$ plane
  that can be obtained by analysing available SDSS measurements of the
  luminosity function ($\Phi$), galaxy-galaxy lensing signal
  ($\Delta\Sigma$) and the
  projected clustering measurements ($w_\rmp$). The 68, 95 and 99 percent
  confidence levels shown by green, yellow and blue contours
  correspond to the constraints possible when only the $\Phi$ and
  $\Delta\Sigma$ data
  are analysed, when only $\Phi$ and $w_\rmp$ data are used, and when all
  three $\Phi$, $\Delta\Sigma$ and $w_\rmp$ data are used in conjunction, respectively. The
  three panels correspond to different priors assumed on the
  cosmological parameters as indicated at the top of each panel. Note
  that, in addition to all the CLF parameters and other secondary
  cosmological parameters, we have also marginalized over the nuisance
  parameters of our model, ${\eta}$ and $\psi$. The dotted
  contours in the middle panel show the 68, 95 and 99 percent
  confidence levels obtained by WMAP7.}
\label{figLCDM}
\end{figure*}
\begin{figure*}
\centerline{\psfig{figure=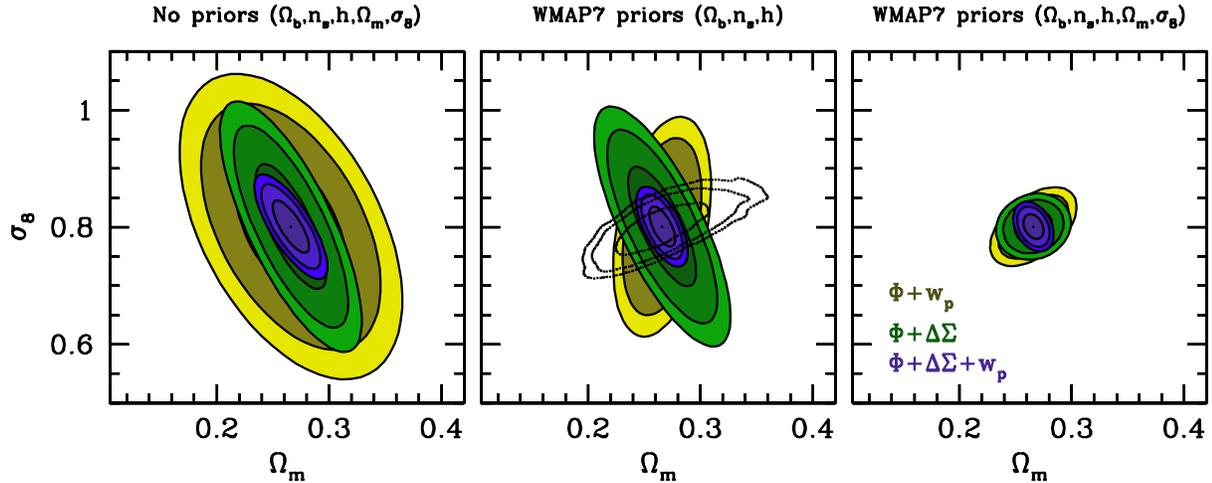,width=0.9\hdsize}}
\caption{ Same as Fig.~\ref{figLCDM}, except for the assumption that
the galaxy-galaxy lensing signal ($\Delta\Sigma$) has been measured upto large
scales ($\sim30\,\mpch$) and is modelled using our analytical
framework.
}
\label{figLCDM_lesd}
\end{figure*}
As is evident from Eq.~(\ref{eq:fish}), the Fisher information matrix
is a (weighted) summation over the derivatives of the model
predictions, $\bolds{\mu}$, with respect to the model parameters,
$\bolds{\theta}$. The constraining power of any given datapoint on a
particular parameter depends on the ratio of the corresponding
derivative to the errorbar on this datapoint. The larger the absolute
value of this ratio, the greater the constraining power.  Since we
assume that the errorbars are a certain fixed percentage of the
observables themselves (see Table~\ref{tab:tab3}), we are interested
in the logarithmic derivatives of the observables with respect to the
model parameters. These derivatives give insight into the power with
which each observable is able to constrain the model parameter of
interest.  The logarithmic derivatives of the luminosity function,
projected galaxy clustering and the galaxy-galaxy lensing signal are
presented in Appendix~A, together with a detailed discussion.

Throughout this paper, we assume that the covariance matrix for the
luminosity function and the galaxy-galaxy lensing signal is diagonal
and calculate the Fisher information matrix by using Equation
\ref{eq:fish}. For the galaxy clustering data, we assume that the
errorbars are correlated in a manner which is quantitatively equal to
the correlations that exist in the measurements of the projected
clustering of SDSS galaxies carried out by \citet{Zehavi2011}. We make
use of Eq.~(\ref{eq:fish2}) to calculate the Fisher information
matrix for the galaxy clustering data.
\begin{figure*}
\centerline{\psfig{figure=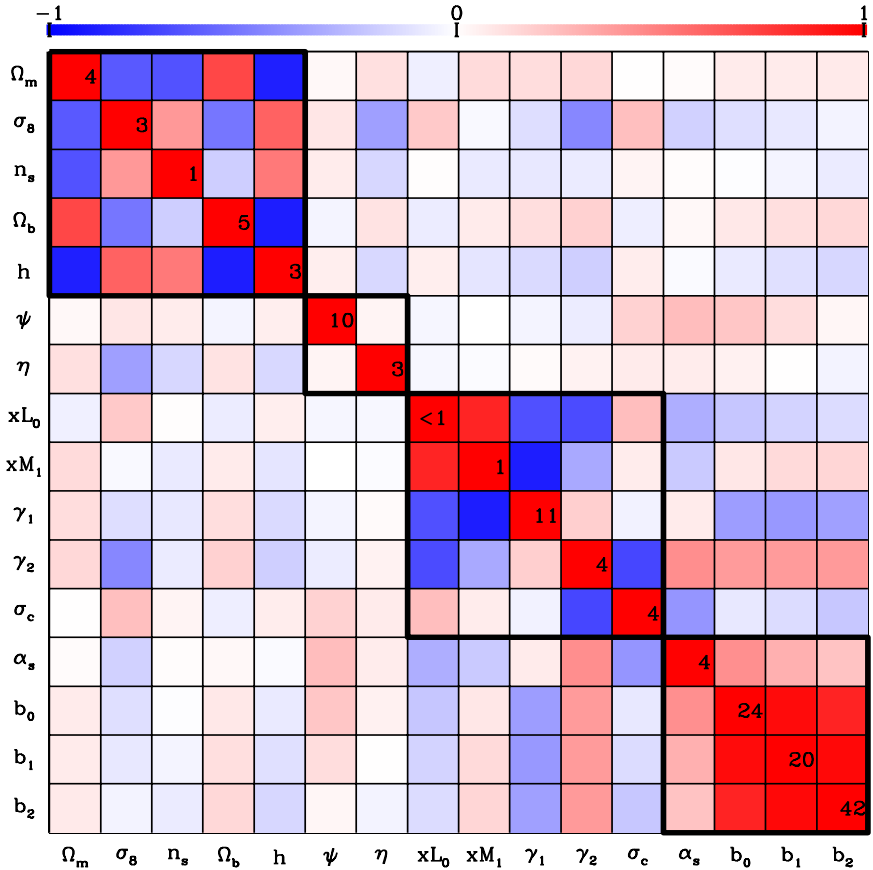,width=0.65\hdsize,height=0.65\hdsize}}
\caption{ Expected cross-correlation coefficients between different
  model parameters when the luminosity function, galaxy clustering and
  galaxy-galaxy lensing signal are analysed together. Red corresponds
  to perfectly correlated constraints on the corresponding parameters,
  while blue corresponds to perfectly anti-correlated constraints. The
  number on the diagonal shows the percentage accuracy with which the
  corresponding parameter can be constrained.}
\label{figCovar}
\end{figure*}
\section{Parameter constraints and covariances}
\label{sec:constraints}

In this section, we forecast the accuracies with which constraints on
cosmological parameters can be obtained given the accuracy of the
current datasets. To obtain these bounds, we first calculate the
Fisher information matrix, $\tilde{F}_{ij}$, by varying the parameters
listed in Tables~\ref{tab:tab1} and \ref{tab:tab2}. We calculate the
Fisher matrix separately for each of the datasets. As each of the
datasets is independently measured, the Fisher information matrix is
additive. The inverse of the Fisher matrix gives the covariance
matrix, ${\cal C}$, and the diagonal elements of this covariance
matrix, ${\cal C}_{ii}$, represents the accuracy with which the $i$-th
parameter can be constrained after marginalizing over the other
parameters. In Appendix~B, we present the procedure we use to obtain
68, 95 and 99 percent confidence ellipses in a plane corresponding to
two parameters. In addition to these constraints, we also forecast the
cross-correlation coefficients, ${\varrho}_{ij}$, that are expected
between the different parameters of our model. We define this
coefficient according to the following equation,
\begin{equation}
{\varrho}_{ij}= \frac{{\cal C}_{ij}}{\sqrt{{\cal C}_{ii}{\cal C}_{jj}}}\,.
\end{equation}

The cross-correlation coefficient is a number that ranges from [-1,1]
and captures the degeneracies inherent in the determination of the
parameters from the dataset. A positive (negative) cross-correlation
coefficient, ${\varrho}_{ij}$, implies that the  $i$-th and $j$-th
parameters are degenerate, such that the effect of increasing one
parameter can be compensated by increasing (decreasing) the other,
after allowing the other parameters to readjust to the change (which
is the essence of marginalization). A value close to zero implies that
the corresponding parameters are only weakly correlated.

\subsection{Model A: Vanilla $\Lambda$CDM}

We first consider the vanilla $\Lambda$CDM model, in which the
Universe is assumed to have a flat geometry, neutrino mass is assumed
to be negligible, the initial power spectrum is assumed to be a single
power-law, and dark energy is modelled as Einstein's cosmological
constant (i.e., has an EoS with parameter $w_0 = -1$). In order to
assess the impact of priors on the cosmological constraints, we perform
three sets of analyses. For the first set, we assume non-informative
priors for the CLF parameters and all of the cosmological parameters.
For the second set, we include prior information about the secondary
cosmological parameters $\Omega_\rmb$, $n_\rms$ and $h$ from the seven
year analysis of the cosmic microwave background data from WMAP
\citep[hereafter WMAP7]{Komatsu2011}. For the third set, we
additionally include priors on the cosmological parameters
$\Omega_\rmm$ and $\sigma_8$ from WMAP7. We denote the second
set of priors to be the fiducial for model A and note that this set
will be used in Paper III to obtain the constraints from actual data.

The constraints on the cosmological parameters $\Omega_\rmm$ and
$\sigma_8$ resulting from the three different sets of priors are shown
in Fig.~\ref{figLCDM} after marginalizing over all CLF parameters, two
nuisance parameters, ${\eta}$ and $\psi$, and the
secondary cosmological parameters, $\Omega_\rmb$, $n_\rms$ and $h$.
The green contours show the 68, 95 and 99 percent confidence regions
that can be obtained using a combination of the luminosity function
and the galaxy-galaxy lensing data. In the absence of prior
information on the secondary cosmological parameters, the two data
sets do not yield particularly tight constraints on $\Omega_\rmm$ and
$\sigma_8$. This is expected in part because the galaxy-galaxy lensing
signal measured from SDSS lacks information from large scales ($r_\rmp
\gta 2 \mpch$). As is evident from the middle panel of
Fig.~\ref{figLCDM}, using prior information on the secondary
cosmological parameters from WMAP7 significantly reduces the
uncertainties on $\Omega_\rmm$ and $\sigma_8$, although a strong
degeneracy of the form $\sigma_8 \propto \Omega_\rmm^{-3.8}$ remains.

\begin{figure}
\centerline{\psfig{figure=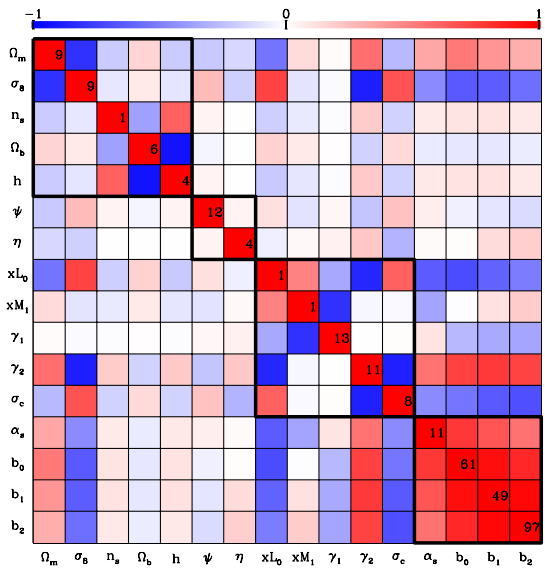,width=0.85\hssize}}
\caption{ Same as Fig.~\ref{figCovar}, except when only the luminosity
function and the excess surface density data are analysed.}
\label{figCovar_LFESD}
\end{figure}
\begin{figure}
\centerline{\psfig{figure=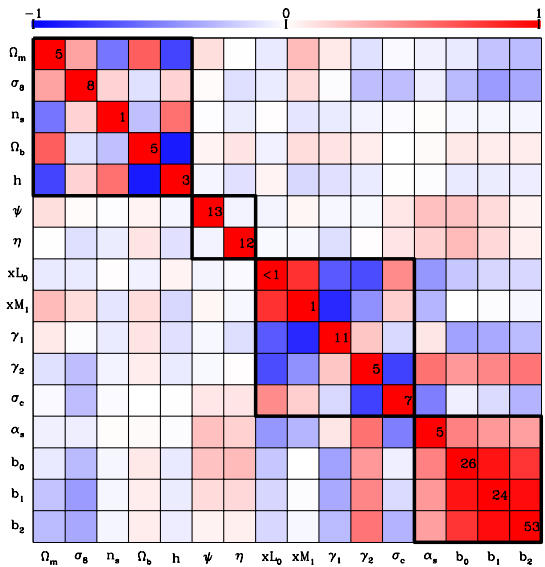,width=0.85\hssize}}
\caption{ Same as Fig.~\ref{figCovar}, except when only the luminosity
function and the projected galaxy clustering data are analysed.}
\label{figCovar_LFWp}
\end{figure}
\begin{figure*}
\centerline{\psfig{figure=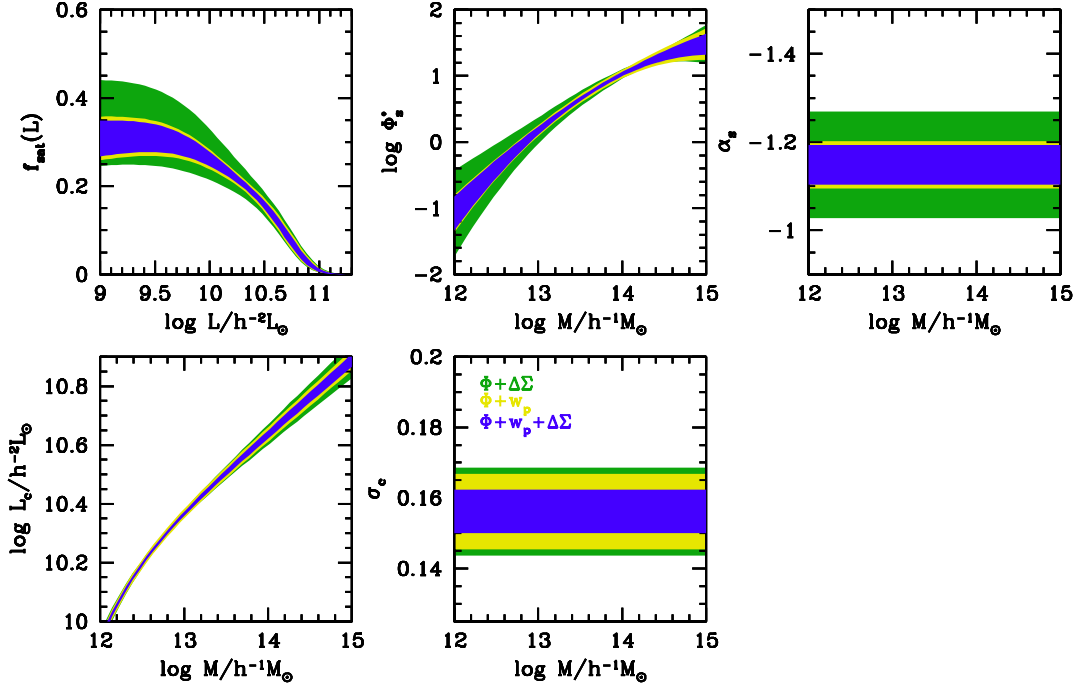,width=0.8\hdsize}}
\caption{ Expected constraints (68 percent confidence) on the
satellite fraction, the satellite CLF normalization, its faint end
slope, the luminosity-mass relation for centrals and the scatter in
this relation using the combination of datasets indicated in the
legend. 
}
\label{fig_clfconst}
\end{figure*}
\begin{figure*}
\centerline{\psfig{figure=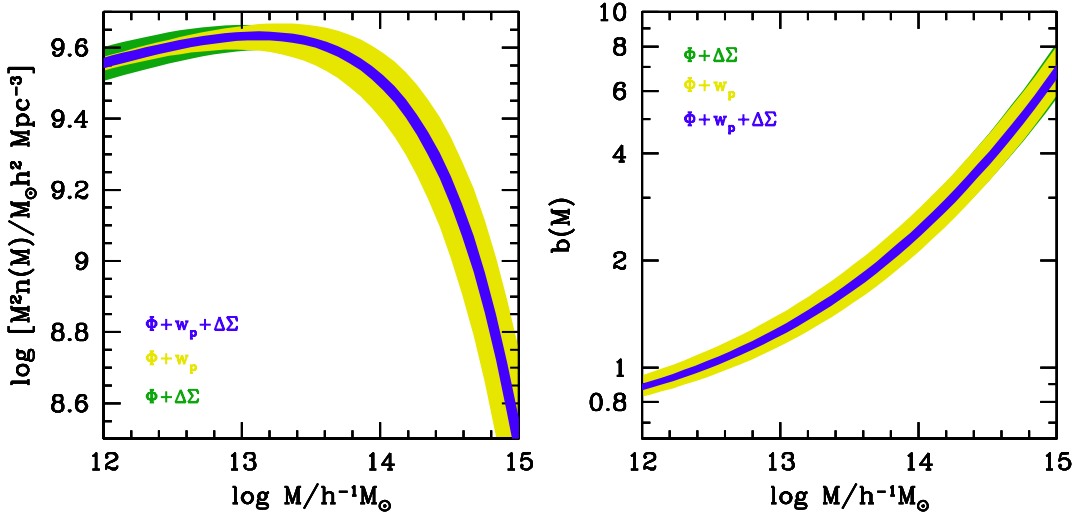,width=0.8\hdsize}}
\caption{ Expected constraints (68 percent confidence) on the mass
function and the bias function using the combinations of datasets
indicated in the legend.
}
\label{fig_cosmoconst}
\end{figure*}
The confidence regions highlighted in  yellow are a result of using a
combination of the luminosity function and the clustering data. The
constraints without the prior information on secondary parameters
(left-hand panel) are significantly better than in the previous case. This
is primarily due to the fact that the clustering data extend out to
larger radii than the lensing data. As shown in Appendix~A, the
clustering data on large scales (where the 2-halo term dominates), has
good constraining power for $\Omega_\rmm$ and $\sigma_8$.  Addition of
the prior information on the secondary cosmological parameters from
WMAP7 further improves the constraints (middle panel of
Fig.~\ref{figLCDM}).  Interestingly, the degeneracies obtained using
the luminosity function and the clustering are in an opposite
direction to the degeneracies obtained using the luminosity function
and the galaxy-galaxy lensing signal, with $\sigma_8 \propto
\Omega_\rmm^{11.2}$. This immediately suggests that using the
luminosity function in combination with both clustering and lensing
will be able to yield even tighter constraints. 

The contours highlighted in blue show the combined constraints
possible when all three observables are analysed together. Even in the
absence of informative priors on the secondary cosmological
parameters, $\Omega_\rmm$ and $\sigma_8$ can already be constrained to
very good accuracy, competitive with current constraints in the
existing literature (see discussion below). In particular, the
constraints on $\Omega_\rmm$ and $\sigma_8$ from a joint model
outperform those obtained by a post-combination of constraints from
each of the dataset analysed separately. A joint analysis is able to
effectively constrain the CLF parameters, which is crucial for teasing
out the tightest possible constraints on the cosmological parameters.
This underscores the importance of having each of these observables
measured from a single uniform survey. For comparison, the dotted
contours in the middle panel of Fig.~\ref{figLCDM} show the
constraints on $\Omega_\rmm$ and $\sigma_8$ from the WMAP7 analysis of
the cosmic microwave background anisotropies.  Note that these
constraints are very comparable (in magnitude) to what is achievable
using our analysis, albeit with roughly orthogonal degeneracies (cf.
blue contours in left-hand panel). When using the WMAP7 priors on the
secondary cosmological parameters, the constraints on $\Omega_\rmm$
and $\sigma_8$ from the combination of luminosity function, clustering
and galaxy-galaxy lensing using present-day data from SDSS become much
tighter than those from WMAP7 alone. Including the WMAP7 priors on
$\Omega_\rmm$ and $\sigma_8$ results in the constraints shown in the
right-hand panel of Fig.~\ref{figLCDM}. Because the degeneracies on
$\Omega_\rmm$ and $\sigma_8$ from WMAP7 are roughly orthogonal to
those resulting from our analysis, the combination results in
extremely tight constraints, even when excluding, for example, the
lensing or clustering data.

\begin{figure*}
\centerline{\psfig{figure=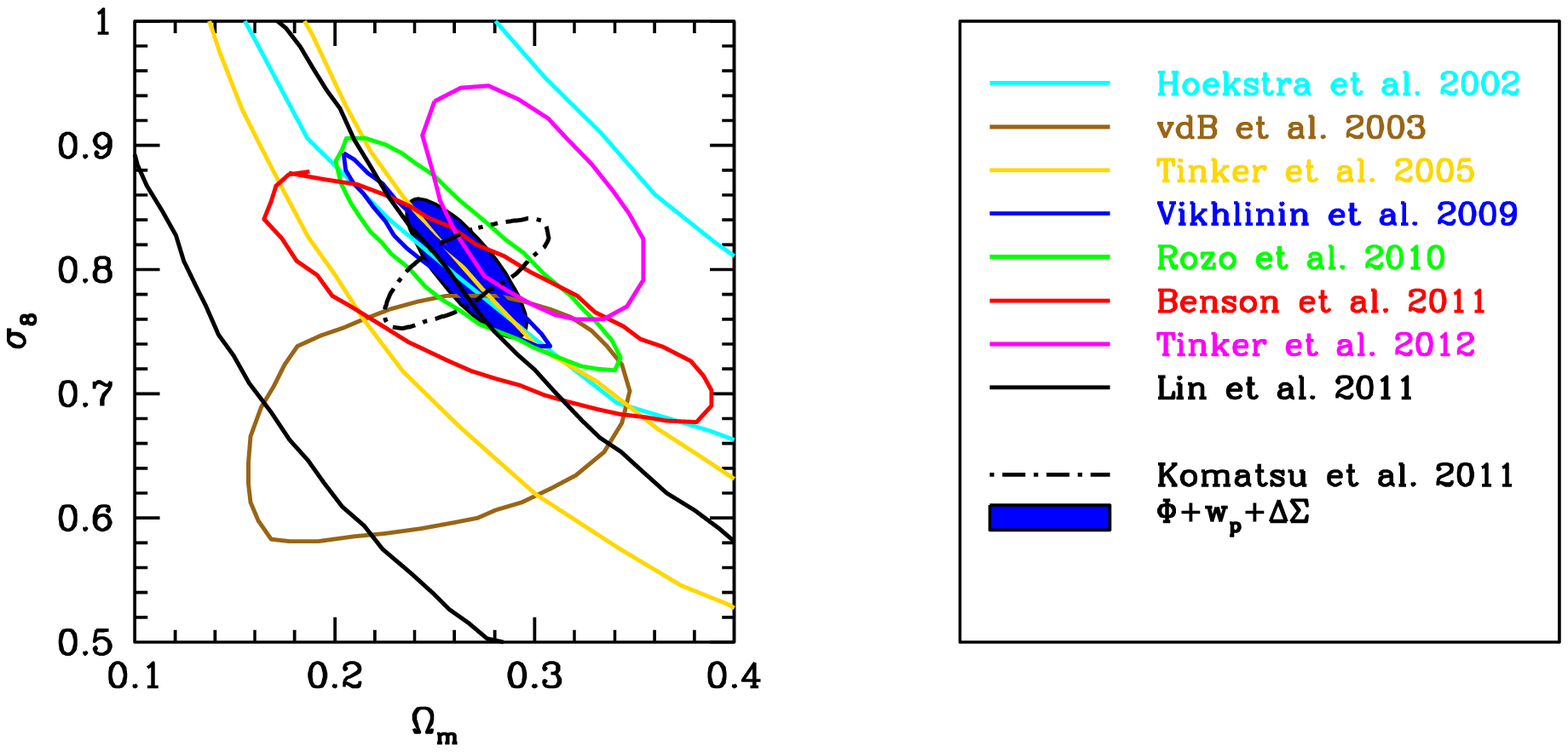,width=0.8\hdsize}}
\caption{ Comparison of the constraints (68 percent confidence) on
    $\Omega_\rmm$ and $\sigma_8$ possible from our joint analysis of
    the luminosity function, galaxy clustering and galaxy-galaxy
    lensing (shown as a blue shaded region and does not include any
    prior information on the secondary cosmological parameters) with
    existing constraints on these parameters from a number of
    independent methods, such as, abundance of massive clusters
    \citep{Vikhlinin2009,Rozo2010,Benson2011}, cosmic shear
    measurement \citep{Hoekstra2002,Lin2011} and halo occupation
    distribution modelling of galaxy clustering
    \citep{vdb2003,Tinker2005,Tinker2012}. The WMAP7 constraints are
    shown using dot-dashed contours.}
\label{figoms8comp}
\end{figure*}
As indicated earlier, the galaxy-galaxy lensing data that we are using
to forecast the cosmological constraints only extend to projected
spatial scales of $2\mpch$. In Fig.~\ref{figLCDM_lesd}, we explore how
the constraints on $\Omega_\rmm$ and $\sigma_8$ would improve if the
galaxy-galaxy lensing was also measured and modelled on scales of
about $30\,\mpch$. Such measurements would be available in the near
future using data from the SDSS (U. Seljak, private communication). A
comparison of these constraints with those shown in
Fig.~\ref{figLCDM_lesd}, shows that the largest improvement
corresponds to the case where we obtain cosmological constraints by
combining the luminosity function and the galaxy-galaxy lensing data
without any priors from WMAP7. However, when using the fiducial set of
priors on the secondary cosmological parameters (or the full set of
priors from WMAP7), the constraints on $\Omega_\rmm$ and $\sigma_8$
are expected to improve only marginally.

In Fig.~\ref{figCovar}, we show the cross-correlation coefficients for
our model parameters expected from a joint analysis of the luminosity
function, galaxy clustering and galaxy-galaxy lensing and using the
fiducial prior set. Squares coloured in red (blue) indicate the
presence of a positive (negative) cross-correlation coefficient.
Squares devoid of colour indicate a cross-correlation coefficient
close to zero. The number on the diagonal squares is the percentage
accuracy (rounded to the nearest integer) with which the corresponding
parameter can be constrained after marginalizing over all other
parameters. The first five parameters are the cosmological parameters,
the next two are the nuisance parameters, followed by the five central
CLF parameters and finally the 4 satellite CLF parameters complete the
set. Of the CLF parameters, x$L_0$, x$M_1$, $\gamma_2$ and
$\sigma_{c}$, which describe the normalizations, the high mass end
slope and the scatter of the $\tilde{L_\rmc}(M)$ relation are the
parameters with constraints that are better than $\sim 4$ percent.  The
familiar degeneracy between x$L_0$ and x$M_1$ is manifested by a
positive cross-correlation coefficient (see Appendix~A). In general,
the central CLF parameters are very tightly coupled with each other
and show a number of degeneracies. The same is true for the satellite
CLF parameters. In particular the parameters $b_0$, $b_1$ and $b_2$
that describe the normalization, $\Phi_*(M)$, are constrained very
poorly and in a highly correlated fashion. There is also a
non-negligible cross-talk between the central CLF parameters,
satellite CLF parameters and the nuisance parameters.

However, this figure clearly shows that the cross-correlation matrix
nicely separates out into a block diagonal form. Most importantly, the
block of the cosmological parameters is found to have only weak
correlations with the blocks of CLF parameters and nuisance
parameters.  This implies that the constraints on the cosmological
parameters are fairly robust to details regarding the modelling of the
halo occupation statistics. This is in contrast with
Fig.~\ref{figCovar_LFESD}, where we show the cross-correlation matrix
expected from the combination of the luminosity function and the
currently existing galaxy-galaxy lensing data. This combination shows
significant degeneracies between the cosmological parameters and a
number of CLF parameters. These degeneracies are suppressed by using
the combination of the luminosity function and the galaxy clustering
data, as can be seen from Fig.~\ref{figCovar_LFWp}. Analyzing all the
three data sets together further improves the constraints on
$\Omega_m$ and $\sigma_8$.

One of the significant cross-correlation between parameters from the
cosmological block and the other parameters, when all the three
datasets are analysed together, is between $\sigma_8$ and the nuisance
parameter ${\eta}$, which characterizes the uncertainty in the
calibration of the normalization of the concentration-mass relation.
This degeneracy is a manifestation of the well-known fact that the
normalization of the concentration-mass relation depends on
$\sigma_8$: dark matter haloes in a universe with a larger value of
$\sigma_8$ collapse earlier, when the universe is denser, and
therefore have higher concentrations \citep[e.g.,][]{Maccio2008}.
Hence, an increase in $\sigma_8$ can be countered by a decrease in
${\eta}$, which explains why the cross correlation coefficient
is negative.

We also expect to obtain excellent constraints on the halo occupation
distribution of galaxies from our analysis.  The upper panels of
Fig.~\ref{fig_clfconst} show the constraints, using the fiducial prior
set, on the satellite fractions as a function of luminosity and the
normalization and faint-end slope of the satellite CLF as a function
of halo mass, respectively, for different combinations of data, as
indicated in the legend. The bottom panels show the constraints on the
average halo mass-luminosity relationship of central galaxies and its
scatter. It is clear that the combination of luminosity function and
clustering outperforms the combination of the luminosity function and
the galaxy-galaxy lensing signal in constraining the halo occupation
distribution parameters, and the constraints improve only marginally
when analysing all three datasets together. The importance of a joint
analysis, however, is very clear from Figure~\ref{fig_cosmoconst},
which shows the expected accuracy with which the halo mass function
and the halo bias function can be constrained. These two quantities
are the primary source of our cosmological information. The
constraints on these quantities are fairly broad when the
galaxy-galaxy lensing data are left out, especially at the high mass
end. However, they improve considerably with the inclusion of the
galaxy-galaxy lensing data.

Finally, in Fig.~\ref{figoms8comp}, we compare the constraints on
$\Omega_\rmm$ and $\sigma_8$ possible from our joint analysis with
existing constraints from other independent studies using a variety of
observations\footnote{The constraints shown in Figure
\ref{figoms8comp} were obtained from the respective manuscripts using
the web application Dexter \citep{Demleitner2001}.}. To avoid
excessive overcrowding in the figure, we only show the 68 percent
confidence regions from each study. As representative of the cosmic
shear studies, we show results from \citet{Hoekstra2002} using cyan
contours and the latest analysis of the SDSS Stripe 82 co-added data
by \citet{Lin2011} using black contours. Clearly, the current data on
cosmic shear measurements is not yet able to put interesting
constraints on $\Omega_\rmm$ and $\sigma_8$. However, this situation
is expected to improve drastically in the near future, thanks to a
number of deep, large-scale photometric surveys planned for this
decade. More stringent constraints on $\Omega_\rmm$ and $\sigma_8$
have come from studies of cluster abundances. As representative of
these studies, we show the constraints obtained by
\citet{Vikhlinin2009} using X-ray clusters from the Chandra Cluster
Cosmology Project (blue contours), by \citet{Rozo2010} using optically
selected MaxBCG clusters (green contours), and by \citet{Benson2011}
using SZ-selected clusters from the South Pole Telescope (red
contours).

On this compilation plot, we also show the constraints obtained by two
studies that used halo occupation modeling to study the abundances and
clustering of galaxies. \citet{vdb2003} used the observed abundance
and luminosity dependence of the correlation length of 2dFGRS
galaxies, combined with independent constraints on the mass-to-light
ratios of galaxy clusters, which resulted in the constraints depicted
by the brown contours.  Using a similar method, \citet{Tinker2005}
modelled the small-scale clustering of SDSS galaxies and used the
mass-to-light ratio on cluster scales to obtain the constraints
depicted in orange.
  
More recently, \citet{Tinker2012} analysed the small scale
($<3\,\mpch$) clustering of galaxies from SDSS and the mass-to-number
ratio of the maxBCG cluster sample \citep[as obtained from the weak
lensing analysis of][]{Sheldon2009} to constrain $\Omega_\rmm$ and
$\sigma_8$.  The 68 percent confidence contours from their analysis
are shown using magenta contours.  When compared to the forecasted
constraints from our analysis (shown as a shaded region and including
no prior information about the secondary cosmological parameters), it
is clear that modelling the entire galaxy-galaxy lensing signal,
rather than just using the mass-to-number ratio, yields additional
information about the cosmological parameters. The fact that the
clustering and lensing samples used by \citet{Tinker2012} are not
well-matched (the clustering data is taken from the spectroscopic
SDSS, whose median redshift is very different from that of the
photometric sample used to create the maxBCG cluster catalog) also
introduces additional systematics which result in the weaker
constraints shown in the figure. 

Note that all these different constraints use different priors on the
secondary cosmological parameters (or none). Therefore their merits
cannot be compared directly. However, this figure does demonstrate
that the joint analysis of the abundances, clustering and lensing of
galaxies, as advocated in this paper, is an extremely powerful way of
constraining the parameters, $\Omega_\rmm$ and $\sigma_8$. Such an
analysis is forecast to yield constraints, even without any additional
priors on the secondary cosmological parameters, that are competitive,
if not better than, any of the previous constraints, including the
WMAP7 data itself.

\subsection{Model B: Massive Neutrinos}

Next we consider cosmological models with massive neutrinos. We
retain the assumption of a flat Universe and the standard
dark energy EoS, $w_0 = -1$. We assume uninformative priors on the CLF
parameters and the primary cosmological parameters of interest,
$\Omega_\rmm$, $\sigma_8$ and $\Omega_{\nu}$. The secondary
cosmological parameters have priors from the 7-year WMAP data
analysis. 
The density parameter $\Omega_{\nu}$ is related to the mass of the
neutrino species, $m_{\nu}$ such that
\begin{equation}
\Omega_{\nu}h^2=\frac{\sum_i m_{\nu,i}}{94 {~\rm eV}} \,.
\end{equation}
We perform the Fisher analysis around the cosmological parameters for
model B displayed in Table~\ref{tab:tab1}. The fiducial value for the
neutrino density parameter is assumed to be $\Omega_\nu=0.004$ which
corresponds to $\sum m_\nu=0.184\,\rme\rmV$.

\begin{figure*}
\centerline{\psfig{figure=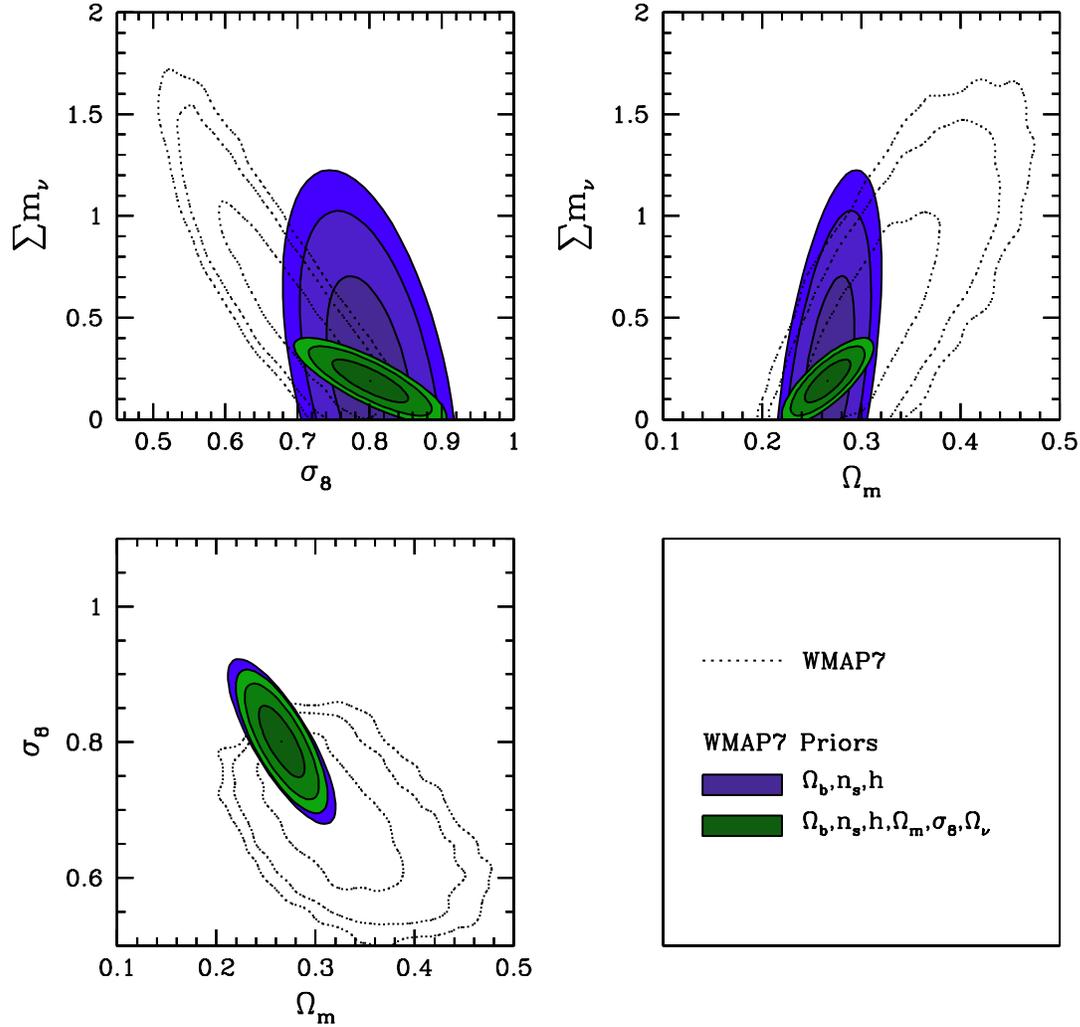,width=0.8\hdsize}}
\caption{ The 68, 95 and 99 percent confidence constraints on the
  parameters $\sum m_\nu$, $\sigma_8$ and $\Omega_\rmm$ after
  marginalizing over the CLF parameters and other nuisance parameters
  in our model for two different priors as indicated in the
  legend. The confidence contours from the WMAP7 analysis are shown
  using dotted lines.}
\label{fignuCDM}
\end{figure*}

The presence of massive neutrinos in the Universe affects the matter
power spectrum primarily by suppressing the density fluctuations on
scales smaller than the neutrino free streaming scale. We use the
framework of \citet{Eisenstein1999} to obtain the linear power
spectrum of matter fluctuations in the presence of massive neutrinos.
\citet{Kiakotou2008} have suggested a minor modification to this
fitting function which leads to a more accurate prediction of the
linear power spectrum when the number of neutrino species with
degenerate masses $N_{\nu}=3$.  We use this modification to get the
linear power spectrum of matter fluctuations. 
When computing the non-linear matter power spectrum, and the halo mass
and bias functions from the linear theory matter power spectrum, we
assume that the calibrations which fit the standard $\Lambda$CDM
simulations also generalize to the cosmological models that include
massive neutrinos.  In particular, we assume that the non-linear power
spectrum is obtained from the linear power spectrum using the HALOFIT
prescription given by \citet{Smith2003} with the modifications suggested
recently by \citet{Bird2012} based upon numerical simulations including massive
neutrinos, and that the dependence of the halo mass and bias functions on the
linear matter power spectrum is given by parameters calibrated by
\citet{Tinker2008}. In addition, we also assume that the density distribution
in halos follows the NFW profile with the concentration-mass relation
given by the calibration obtained by
\citet{Maccio2007}.

The constraints on the parameters $\Omega_\rmm$, $\sigma_8$ and the
sum of massive neutrinos that can be obtained from a joint analysis of
the abundance, clustering and weak lensing of galaxies are shown in
Fig.~\ref{fignuCDM}. The blue contours represent constraints when
using the fiducial set of priors on the secondary cosmological
parameters.  As before, we have marginalized over all the CLF
parameters, the nuisance parameters and the other cosmological
parameters. The addition of the parameter, $\sum m_{\nu}$, as
expected, leads to more freedom for $\Omega_\rmm$ and $\sigma_8$ and
this slightly weakens the constraints in the $\Omega_\rmm-\sigma_8$
plane compared to model A.  The expected error on the sum of neutrinos
is not small enough to rule out the massless neutrino case (if at
least, as assumed here, the true sum of neutrino masses is $\sum
m_{\nu}=0.184\,\eV$). The dotted contours show the results from the
analysis of the WMAP7 data. It is clear that both analyses on their own
allow values for the sum of neutrino masses as high as 1 eV at 95
percent confidence.  However, the cosmic microwave background data, in
such a scenario, requires a very low value for $\sigma_8\,\sim\,0.6$.
Our analysis can rule out such values of $\sigma_8$ at very high
confidence. This shows that combining the constraints from our
analysis with the cosmic microwave background results can certainly
provide a significantly better upper limit on the sum of neutrino
masses. This, in essence, is very similar to how the addition of cluster
abundances constraints improves the constraints on neutrino masses
from the WMAP7 analysis \citep[e.g.,][]{Benson2011}. 

The green confidence contours shown in Fig.~\ref{fignuCDM} represent
the constraints on $\Omega_\rmm$, $\sigma_8$ and $\sum m_{\nu}$
possible from our analysis when using all prior information
available from WMAP7. If the sum of neutrino masses is truly as large
as we have assumed in our analysis, such a combination will allow the
tantalising possibility of the first signs of detection of a non-zero
sum of neutrino masses at $\sim 3\sigma$ confidence.
\begin{figure*}
\centerline{\psfig{figure=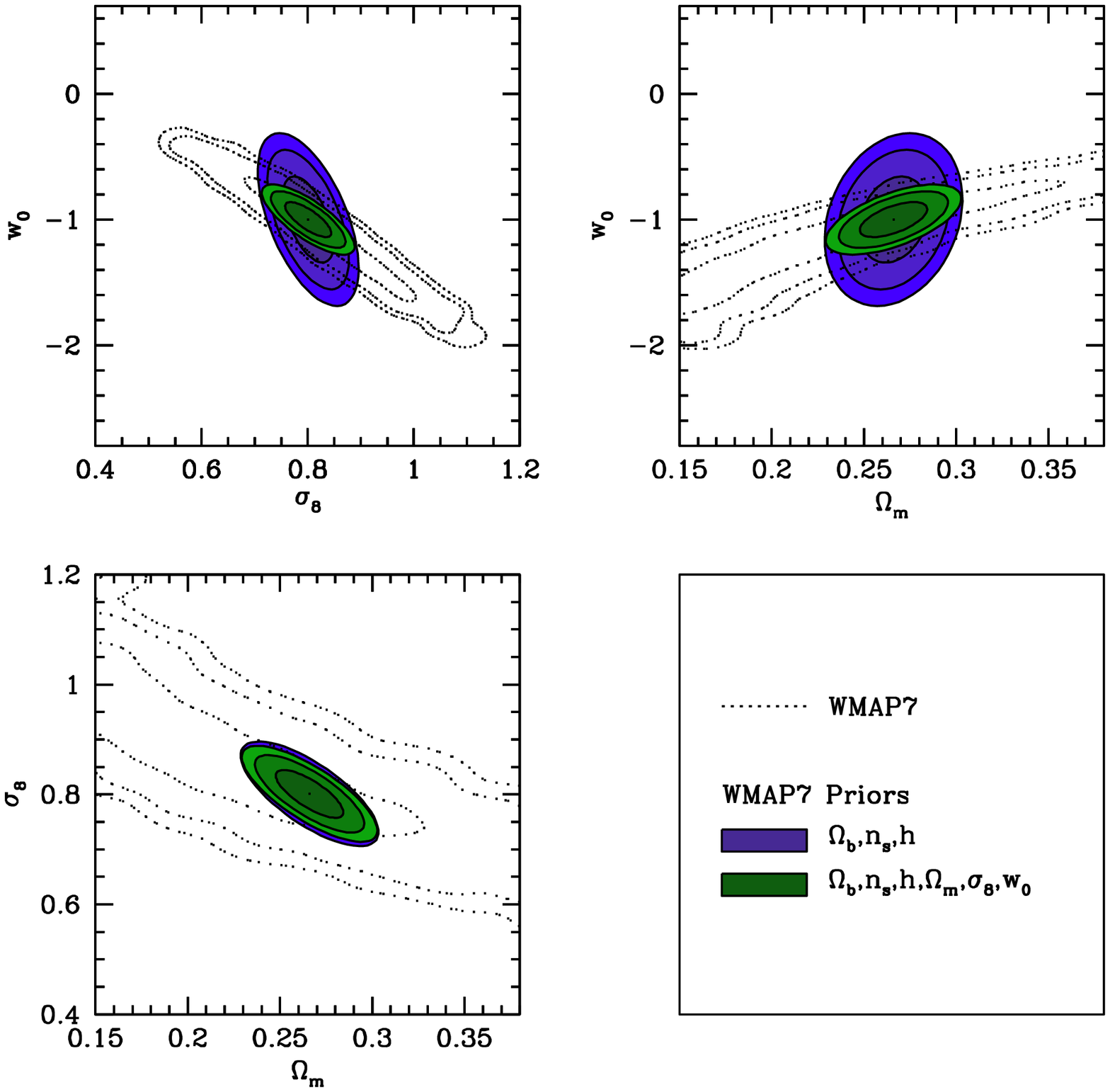,width=0.8\hdsize}}
\caption{ The 68, 95 and 99 percent confidence constraints on the
parameters $w_0$, $\sigma_8$, $\Omega_\rmm$ from our analysis after
marginalizing over the CLF parameters and other nuisance parameters in
our model for two different priors as indicated in the legend. The
confidence contours from the WMAP7 analysis are shown using dotted
lines.
  }
\label{figwCDM}
\end{figure*}

\subsection{Model C: Dark Energy Equation of State}

Finally, we focus on models with a modified EoS for the dark energy,
i.e., with $w_0 \ne 1$. For these models, we restore the assumption of
massless neutrinos and we maintain the assumption of a flat
Universe. Modifications to the dark energy EoS change the expansion
history of the Universe and the growth rate of structure
formation. The expansion factor $E(z)$ is given by
\begin{equation}
E(z)=[\Omega_\rmm\,(1+z)^3+\Omega_{\Lambda}\,(1+z)^{3(1+w_0)}
]^{1/2}\,,
\end{equation}
whereas the dependence of the growth factor on redshift can be
calculated by solving the second order differential equation obtained
by combining the continuity and Euler equations in the linear regime
(see \citealt{Mo2010}),
\begin{equation}
\ddot{g}+2\,H\,\dot{g} = \frac{3}{2} \Omega_\rmm\,H_0^2\,(1+z)^3\,g\,.
\end{equation}
Changing the variables from time to $\eta=\ln a(t)$, this equation can
be written as
\begin{equation}
    g'' + \left[ 2 + \frac{d\ln E}{d\eta}\right] g' =
    \frac{3}{2E^2(\eta)} \Omega_\rmm (1+z)^3 \,,
\end{equation}
where a dash indicates a derivative with respect to $\eta$. We use
the following two boundary conditions to solve for the growth factor:
(a) $g(a=0.001)=1$ and (b) $g'(a=0.001)=1$, consistent with the
expectation that $g\propto a$ in the matter-dominated era. We
renormalize the scale factor to equal unity at redshift $z=0$, once
the solution has been found.

These changes to the growth factor that arise from having the dark
energy EoS differ from $w_0=-1$ propagate in the redshift evolution of
the linear matter power spectrum, and thereby affect the halo mass
function, the halo bias function and the mass dependence of the
concentration parameter that describes the density profile of dark
matter haloes. These changes cause the galaxy abundance, the galaxy
clustering and the galaxy-galaxy lensing signal to deviate from the
`standard' vanilla-$\Lambda$CDM predictions. We once again assume that
the calibrations required to compute the non-linear matter power
spectrum, the halo mass function, the halo bias function and the halo
concentrations from the linear theory power spectrum are the same as
in the standard $\Lambda$CDM case.

The constraints on the dark energy EoS and the cosmological parameters
$\Omega_\rmm$ and $\sigma_8$ that are achievable from our joint
analysis, using the WMAP7 priors on the secondary cosmological
parameters, are shown in Fig.~\ref{figwCDM}. The dotted lines
represent the 68, 95 and 99 percent confidence intervals from the
WMAP7 data, and are shown for comparison.  Note that our analysis
results in degeneracies that are very similar to those from the WMAP7
analysis, whereby an increase in $\sigma_8$ can compensate a decrease
in the value of $w_0$. The slight differences in the degeneracy
directions, however, allow tighter constraints when combining both
data sets. The resulting confidence intervals are shown using green
contours.

\begin{figure*}
\centerline{\psfig{figure=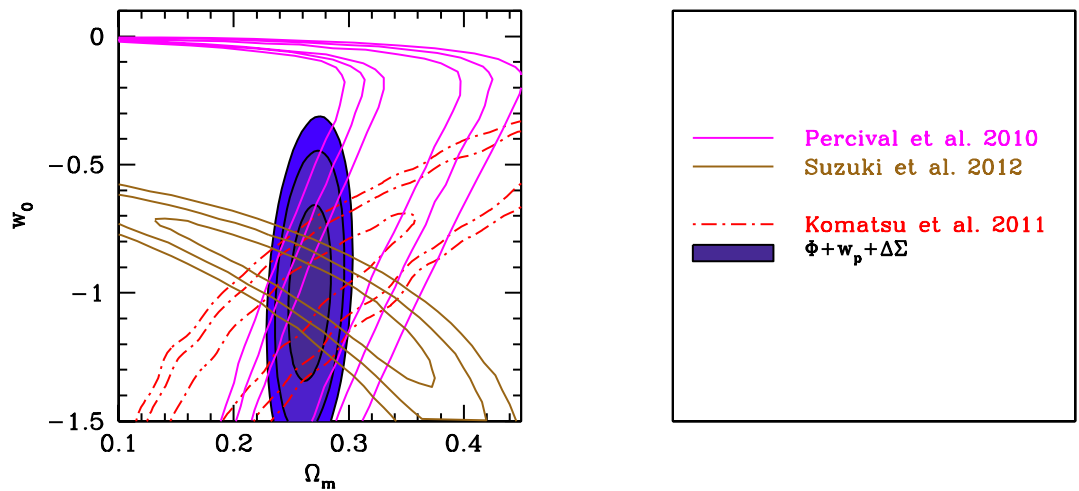,width=0.8\hdsize}}
\caption{ Comparison of the constraints (68, 95 and 99 percent
  confidence) on $\Omega_\rmm$ and $w_0$ possible from our joint
  analysis of the luminosity function, galaxy clustering and
  galaxy-galaxy lensing (shown as a shaded region and includes
  WMAP7 priors on secondary cosmological parameters) with existing
  constraints on these parameters.  The magenta contours are from the
  analysis of the baryon acoustic feature from SDSS data by
  \citet{Percival2010}, the brown contours are from the analysis of
  SNeIa data by \citet{Suzuki2012} and the red contours are from the
  analysis of WMAP7 data by \citet{Komatsu2011}.  }
\label{figomw0comp}
\end{figure*}

In Fig.~\ref{figomw0comp}, we compare the constraints on $\Omega_\rmm$
and $w_0$ that can be obtained from our joint analysis with those
obtained from other cosmological probes. Supernovae of Type Ia (SNeIa)
and measurements of the baryon acoustic feature (BAF) in the
clustering of galaxies can be used as standard candles and standard
rulers, respectively, in order to map the expansion history of the
Universe as a function of redshift. Brown contours indicate the 68, 95
and 99 percent confidence intervals on the parameters $\Omega_\rmm$
and $w_0$ obtained by \cite{Suzuki2012} from the Union2.1 compilation
of SNeIa. The constraints obtained by \citet{Percival2010}, who
measured the BAF by combining SDSS and 2dFGRS data are shown using
magenta contours, while the constraints from the analysis of the WMAP7
data are shown using red contours. Each of these constraints have
severe degeneracies in the $\Omega_\rmm-w_0$ plane, but are very
powerful when used in conjunction. The constraints forecast for our
analysis, shown using filled blue contours, indicates that our results
will be degenerate in yet another direction and can therefore
significantly add to our knowledge of the cosmological parameters
$\Omega_\rmm$ and $w_0$. It is worth mentioning that unlike our
analysis, the probes of expansion history such as SNe1a are
insensitive to the parameter $\sigma_8$. The overall amplitude of the
clustering measurements used to detect the BAF does depend upon
$\sigma_8$, but cannot be used to constrain it, due to its degeneracy
with the value of galaxy bias. Our analysis, on the other hand, is
able to constrain both $\sigma_8$ and galaxy bias simultaneously due
to our modelling of small scale clustering and the galaxy-galaxy
lensing signal \citep[see also][]{Seljak2005}.

\section{Summary} 
\label{sec:summary}

The distribution of galaxies, which is routinely mapped out by large
scale galaxy surveys, contains a wealth of information about the
cosmological parameters. Unfortunately, harvesting this information is
not straightforward because galaxies are biased tracers of the matter
distribution. We have presented a powerful method that can be used to
simultaneously solve for both the cosmological parameters as well as
the galaxy bias.  Our method relies on an accurate halo occupation
distribution modelling of galaxy abundances, galaxy clustering and
galaxy-galaxy lensing, which probes scales well into the non-linear
regime of structure formation.

In Paper I we developed an analytical framework to predict each of
these observables, given a cosmology and the CLF parameters that
describe the halo occupation statistics of galaxies. In this paper, we
have presented the strength of each of these
observables (see Appendix~A) to constrain our model parameters. We have also performed a Fisher matrix
analysis for a standard $\Lambda$CDM cosmology with flat geometry. We
primarily focused on the cosmological parameters $\Omega_\rmm$ and
$\sigma_8$, and treated the other cosmological parameters
($\Omega_\rmb$, $n_\rms$ and $h$) as secondary parameters, for which
we adopted priors taken from the seven year analysis of the WMAP data.

We have shown that an analysis of the luminosity function of galaxies
combined with galaxy-galaxy lensing is marred by a number of
degeneracies between the CLF parameters and the cosmological
parameters, which effectively weakens the cosmological constraints.
However, when this analysis is complemented with clustering data, a
number of these degeneracies are broken.  For the standard
$\Lambda$CDM cosmology, we forecast that the constraints on
$\Omega_\rmm$ and $\sigma_8$ that are achievable from our joint
analysis, using data that is already available from the SDSS, are of
the order of $3$ and $2$ percent, respectively. Such constraints will
be competitive with (and complementary to) the tightest constraints
that are presently available (see Fig.~\ref{figoms8comp}). We will
perform such an analysis in Paper~III (Cacciato et al, in
preparation).

One of the most important results of this paper is the demonstration
that the covariance of the posterior distribution of parameters from our joint analysis of the luminosity
function, galaxy clustering and galaxy-galaxy lensing has a block
diagonal form (see Fig.~\ref{figCovar}). In particular, the
cosmological parameters form a separate block, which is only weakly
correlated with the other blocks (2 nuisance parameters, 5 parameters
that describe the CLF of central galaxies, and 4 parameters that
describe the CLF of satellites). This implies that the cosmological
constraints will be robust to uncertainties related to details of the
halo occupation statistics, and that we can simultaneously obtain
tight constraints on the halo occupation statistics, properly
marginalized over the uncertainties in the cosmological parameters.

We have also investigated two extensions of the standard $\Lambda$CDM
model. In the first extension we included massive neutrinos in our
analysis. We showed that although the constraints on the sum of the
neutrino masses, $\sum m_\nu$, from our analysis are predicted to be
of the order of $1$ eV (at $95$ percent confidence, assuming realistic
errors on the data), the degeneracies between $\sum m_\nu$ and
$\sigma_8$ are oriented differently than those from the WMAP7
analysis. Consequently, the combination of our results with those from
WMAP7 can significantly improve the constraints on $\sum m_\nu$.  In
particular, we have shown that such an analysis should be able to
provide constraints of the order of 0.2 eV at 99 percent confidence.

As a second extension of $\Lambda$CDM, we also allowed for a
non-standard EoS for the dark energy, characterized by $w_0 \ne -1$.
We showed that the current data on the abundances, clustering and
lensing of galaxies in SDSS is already sufficient to put interesting
constraints on $w_0$, especially when combined with the WMAP7 data.
In particular, the constraints on $w_0$ from our analysis are forecast
to correspond to $\Delta w_0=0.5$ (95 percent confidence, with priors
on the secondary cosmological parameters). This further improves to
$\Delta w_0=0.2$ when using the full prior information available from
WMAP7 (see Fig.~\ref{figwCDM}).  Since the SDSS data considered here
only spans a very narrow range in redshift ($z\lsim0.2$), there is a
huge potential for improvement by measuring the abundances, clustering
and lensing of galaxies at higher redshifts.

Finally, we note that although the main focus of our joint analysis is
to constrain cosmological parameters, we have also shown that we will
be able to obtain a detailed, statistical description of the
galaxy-matter connection, as parameterized by the CLF, fully
marginalized over uncertainties in the cosmological parameters.  In
particular, our analysis can provide excellent constraints on the
average relation between halo mass and central galaxy luminosity,
including its scatter, on the satellite fraction as function of
luminosity, and on the conditional luminosity function of satellites.
These constraints characterize, among others, the efficiency with
which haloes of different masses are able to convert their
cosmological share of baryons into stars (i.e., galaxies).  This will
be of great value to inform models of galaxy formation and evolution.
Another powerful application of a tightly constrained CLF is the
construction of realistic mock galaxy catalogues (see Paper~I for an
example). The galaxies in these mock catalogues will, by construction,
have abundances, clustering and a cross-correlation with matter,
that are all consistent with current data.

\section{Acknowledgments} 

We are grateful to Matthew Becker, Wayne Hu, Alexie
Leauthaud, Yin Li,
Eduardo Rozo, Jeremy Tinker and Martin White for many interesting
discussions and possible extensions of the research described in this
paper. SM acknowledges support from the Kavli Institute for
Cosmological Physics at the University of Chicago through the NSF
grant PHY-0551142 and an endowment from the Kavli Foundation.  
The analysis presented in this work has been performed on the Joint
Fermilab - KICP Supercomputing Cluster, supported by grants from
Fermilab, Kavli Institute for Cosmological Physics, and the University
of Chicago. FvdB acknowledges support from the Lady Davis Foundation
for a Visiting Professorship at Hebrew University. FvdB was also
supported in part by the National Science Foundation under Grant No.
NSF PHY11-25915.

\bibliographystyle{aa}
\bibliography{Library}

\begin{thebibliography}{78}
\expandafter\ifx\csname natexlab\endcsname\relax\def\natexlab#1{#1}\fi

\bibitem[{{Abazajian} {et~al.}(2009){Abazajian}, {Adelman-McCarthy},
  {Ag{\"u}eros}, {Allam}, {Allende Prieto}, {An}, {Anderson}, {Anderson},
  {Annis}, {Bahcall}, \& et~al.}]{Abazajian2009}
{Abazajian}, K.~N., {Adelman-McCarthy}, J.~K., {Ag{\"u}eros}, M.~A., {et~al.}
  2009, \apjs, 182, 543

\bibitem[{{Anderson} {et~al.}(2012){Anderson}, {Aubourg}, {Bailey}, {Bizyaev},
  {Blanton}, {Bolton}, {Brinkmann}, {Brownstein}, {Burden}, {Cuesta}, {da
  Costa}, {Dawson}, {de Putter}, {Eisenstein}, {Gunn}, {Guo}, {Hamilton},
  {Harding}, {Ho}, {Honscheid}, {Kazin}, {Kirkby}, {Kneib}, {Labatie},
  {Loomis}, {Lupton}, {Malanushenko}, {Malanushenko}, {Mandelbaum}, {Manera},
  {Maraston}, {McBride}, {Mehta}, {Mena}, {Montesano}, {Muna}, {Nichol},
  {Nuza}, {Olmstead}, {Oravetz}, {Padmanabhan}, {Palanque-Delabrouille}, {Pan},
  {Parejko}, {Paris}, {Percival}, {Petitjean}, {Prada}, {Reid}, {Roe}, {Ross},
  {Ross}, {Samushia}, {Sanchez}, {Schneider}, {Scoccola}, {Seo}, {Sheldon},
  {Simmons}, {Skibba}, {Strauss}, {Swanson}, {Thomas}, {Tinker}, {Tojeiro},
  {Vargas Magana}, {Verde}, {Wagner}, {Wake}, {Weaver}, {Weinberg}, {White},
  {Xu}, {Yeche}, {Zehavi}, \& {Zhao}}]{Anderson2012}
{Anderson}, L., {Aubourg}, E., {Bailey}, S., {et~al.} 2012, ArXiv e-prints

\bibitem[{{Baldauf} {et~al.}(2010){Baldauf}, {Smith}, {Seljak}, \&
  {Mandelbaum}}]{Baldauf2010}
{Baldauf}, T., {Smith}, R.~E., {Seljak}, U., \& {Mandelbaum}, R. 2010, \prd,
  81, 063531

\bibitem[{{Baldry} {et~al.}(2008){Baldry}, {Glazebrook}, \&
  {Driver}}]{Baldry2008}
{Baldry}, I.~K., {Glazebrook}, K., \& {Driver}, S.~P. 2008, \mnras, 388, 945

\bibitem[{{Benson} {et~al.}(2011){Benson}, {de Haan}, {Dudley}, {Reichardt},
  {Aird}, {Andersson}, {Armstrong}, {Bautz}, {Bayliss}, {Bazin}, {Bleem},
  {Brodwin}, {Carlstrom}, {Chang}, {Cho}, {Clocchiatti}, {Crawford}, {Crites},
  {Desai}, {Dobbs}, {Foley}, {Forman}, {George}, {Gladders}, {Halverson},
  {High}, {Holder}, {Holzapfel}, {Hoover}, {Hrubes}, {Jones}, {Joy}, {Keisler},
  {Knox}, {Lee}, {Leitch}, {Liu}, {Lueker}, {Luong-Van}, {Mantz}, {Marrone},
  {McDonald}, {McMahon}, {Mehl}, {Meyer}, {Mocanu}, {Mohr}, {Montroy},
  {Murray}, {Natoli}, {Padin}, {Plagge}, {Pryke}, {Rest}, {Ruel}, {Ruhl},
  {Saliwanchik}, {Saro}, {Schaffer}, {Shaw}, {Shirokoff}, {Song}, {Spieler},
  {Stalder}, {Staniszewski}, {Stark}, {Story}, {Stubbs}, {Suhada}, {van
  Engelen}, {Vanderlinde}, {Vieira}, {Vikhlinin}, {Williamson}, {Zahn}, \&
  {Zenteno}}]{Benson2011}
{Benson}, B.~A., {de Haan}, T., {Dudley}, J.~P., {et~al.} 2011, ArXiv e-prints

\bibitem[{{Berlind} \& {Weinberg}(2002)}]{Berlind2002}
{Berlind}, A.~A. \& {Weinberg}, D.~H. 2002, \apj, 575, 587

\bibitem[{{Bird} {et~al.}(2012){Bird}, {Viel}, \& {Haehnelt}}]{Bird2012}
{Bird}, S., {Viel}, M., \& {Haehnelt}, M.~G. 2012, \mnras, 420, 2551

\bibitem[{{Blake} {et~al.}(2011){Blake}, {Davis}, {Poole}, {Parkinson},
  {Brough}, {Colless}, {Contreras}, {Couch}, {Croom}, {Drinkwater}, {Forster},
  {Gilbank}, {Gladders}, {Glazebrook}, {Jelliffe}, {Jurek}, {Li}, {Madore},
  {Martin}, {Pimbblet}, {Pracy}, {Sharp}, {Wisnioski}, {Woods}, {Wyder}, \&
  {Yee}}]{Blake2011}
{Blake}, C., {Davis}, T., {Poole}, G.~B., {et~al.} 2011, \mnras, 415, 2892

\bibitem[{{Blanton} {et~al.}(2003{\natexlab{a}}){Blanton}, {Brinkmann},
  {Csabai}, {Doi}, {Eisenstein}, {Fukugita}, {Gunn}, {Hogg}, \&
  {Schlegel}}]{Blanton2003a}
{Blanton}, M.~R., {Brinkmann}, J., {Csabai}, I., {et~al.} 2003{\natexlab{a}},
  \aj, 125, 2348

\bibitem[{{Blanton} {et~al.}(2003{\natexlab{b}}){Blanton}, {Hogg}, {Bahcall},
  {Brinkmann}, {Britton}, {Connolly}, {Csabai}, {Fukugita}, {Loveday},
  {Meiksin}, {Munn}, {Nichol}, {Okamura}, {Quinn}, {Schneider}, {Shimasaku},
  {Strauss}, {Tegmark}, {Vogeley}, \& {Weinberg}}]{Blanton2003b}
{Blanton}, M.~R., {Hogg}, D.~W., {Bahcall}, N.~A., {et~al.} 2003{\natexlab{b}},
  \apj, 592, 819

\bibitem[{{Cacciato} {et~al.}(2009){Cacciato}, {van den Bosch}, {More}, {Li},
  {Mo}, \& {Yang}}]{Cacciato2009}
{Cacciato}, M., {van den Bosch}, F.~C., {More}, S., {et~al.} 2009, \mnras, 394,
  929

\bibitem[{{Cole} {et~al.}(2001){Cole}, {Norberg}, {Baugh}, {Frenk},
  {Bland-Hawthorn}, {Bridges}, {Cannon}, {Colless}, {Collins}, {Couch},
  {Cross}, {Dalton}, {De Propris}, {Driver}, {Efstathiou}, {Ellis},
  {Glazebrook}, {Jackson}, {Lahav}, {Lewis}, {Lumsden}, {Maddox}, {Madgwick},
  {Peacock}, {Peterson}, {Sutherland}, \& {Taylor}}]{Cole2001}
{Cole}, S., {Norberg}, P., {Baugh}, C.~M., {et~al.} 2001, \mnras, 326, 255

\bibitem[{{Colless} {et~al.}(2001){Colless}, {Dalton}, {Maddox}, {Sutherland},
  {Norberg}, {Cole}, {Bland-Hawthorn}, {Bridges}, {Cannon}, {Collins}, {Couch},
  {Cross}, {Deeley}, {De Propris}, {Driver}, {Efstathiou}, {Ellis}, {Frenk},
  {Glazebrook}, {Jackson}, {Lahav}, {Lewis}, {Lumsden}, {Madgwick}, {Peacock},
  {Peterson}, {Price}, {Seaborne}, \& {Taylor}}]{Colless2001}
{Colless}, M., {Dalton}, G., {Maddox}, S., {et~al.} 2001, \mnras, 328, 1039

\bibitem[{{Coupon} {et~al.}(2011){Coupon}, {Kilbinger}, {McCracken}, {Ilbert},
  {Arnouts}, {Mellier}, {Abbas}, {de la Torre}, {Goranova}, {Hudelot}, {Kneib},
  \& {Lefevre}}]{Coupon2011}
{Coupon}, J., {Kilbinger}, M., {McCracken}, H.~J., {et~al.} 2011, ArXiv
  e-prints

\bibitem[{{Demleitner} {et~al.}(2001){Demleitner}, {Accomazzi}, {Eichhorn},
  {Grant}, {Kurtz}, \& {Murray}}]{Demleitner2001}
{Demleitner}, M., {Accomazzi}, A., {Eichhorn}, G., {et~al.} 2001, in
  Astronomical Society of the Pacific Conference Series, Vol. 238, Astronomical
  Data Analysis Software and Systems X, ed. F.~R. {Harnden}, Jr., F.~A.
  {Primini}, \& H.~E. {Payne}, 321

\bibitem[{{Dunkley} {et~al.}(2011){Dunkley}, {Hlozek}, {Sievers}, {Acquaviva},
  {Ade}, {Aguirre}, {Amiri}, {Appel}, {Barrientos}, {Battistelli}, {Bond},
  {Brown}, {Burger}, {Chervenak}, {Das}, {Devlin}, {Dicker}, {Bertrand
  Doriese}, {D{\"u}nner}, {Essinger-Hileman}, {Fisher}, {Fowler}, {Hajian},
  {Halpern}, {Hasselfield}, {Hern{\'a}ndez-Monteagudo}, {Hilton}, {Hilton},
  {Hincks}, {Huffenberger}, {Hughes}, {Hughes}, {Infante}, {Irwin}, {Juin},
  {Kaul}, {Klein}, {Kosowsky}, {Lau}, {Limon}, {Lin}, {Lupton}, {Marriage},
  {Marsden}, {Mauskopf}, {Menanteau}, {Moodley}, {Moseley}, {Netterfield},
  {Niemack}, {Nolta}, {Page}, {Parker}, {Partridge}, {Reid}, {Sehgal},
  {Sherwin}, {Spergel}, {Staggs}, {Swetz}, {Switzer}, {Thornton}, {Trac},
  {Tucker}, {Warne}, {Wollack}, \& {Zhao}}]{Dunkley2011}
{Dunkley}, J., {Hlozek}, R., {Sievers}, J., {et~al.} 2011, \apj, 739, 52

\bibitem[{{Eisenstein} \& {Hu}(1999)}]{Eisenstein1999}
{Eisenstein}, D.~J. \& {Hu}, W. 1999, \apj, 511, 5

\bibitem[{{Eisenstein} {et~al.}(2005){Eisenstein}, {Zehavi}, {Hogg},
  {Scoccimarro}, {Blanton}, {Nichol}, {Scranton}, {Seo}, {Tegmark}, {Zheng},
  {Anderson}, {Annis}, {Bahcall}, {Brinkmann}, {Burles}, {Castander},
  {Connolly}, {Csabai}, {Doi}, {Fukugita}, {Frieman}, {Glazebrook}, {Gunn},
  {Hendry}, {Hennessy}, {Ivezi{\'c}}, {Kent}, {Knapp}, {Lin}, {Loh}, {Lupton},
  {Margon}, {McKay}, {Meiksin}, {Munn}, {Pope}, {Richmond}, {Schlegel},
  {Schneider}, {Shimasaku}, {Stoughton}, {Strauss}, {SubbaRao}, {Szalay},
  {Szapudi}, {Tucker}, {Yanny}, \& {York}}]{Eisenstein2005}
{Eisenstein}, D.~J., {Zehavi}, I., {Hogg}, D.~W., {et~al.} 2005, \apj, 633, 560

\bibitem[{{Guy} {et~al.}(2010){Guy}, {Sullivan}, {Conley}, {Regnault},
  {Astier}, {Balland}, {Basa}, {Carlberg}, {Fouchez}, {Hardin}, {Hook},
  {Howell}, {Pain}, {Palanque-Delabrouille}, {Perrett}, {Pritchet}, {Rich},
  {Ruhlmann-Kleider}, {Balam}, {Baumont}, {Ellis}, {Fabbro}, {Fakhouri},
  {Fourmanoit}, {Gonz{\'a}lez-Gait{\'a}n}, {Graham}, {Hsiao}, {Kronborg},
  {Lidman}, {Mourao}, {Perlmutter}, {Ripoche}, {Suzuki}, \& {Walker}}]{Guy2010}
{Guy}, J., {Sullivan}, M., {Conley}, A., {et~al.} 2010, \aap, 523, A7

\bibitem[{{Hoekstra} {et~al.}(2002){Hoekstra}, {Yee}, \&
  {Gladders}}]{Hoekstra2002}
{Hoekstra}, H., {Yee}, H.~K.~C., \& {Gladders}, M.~D. 2002, \apj, 577, 595

\bibitem[{{Huff} {et~al.}(2011){Huff}, {Eifler}, {Hirata}, {Mandelbaum},
  {Schlegel}, \& {Seljak}}]{Huff2011}
{Huff}, E.~M., {Eifler}, T., {Hirata}, C.~M., {et~al.} 2011, ArXiv e-prints

\bibitem[{{Jing} {et~al.}(1998){Jing}, {Mo}, \& {Boerner}}]{Jing1998}
{Jing}, Y.~P., {Mo}, H.~J., \& {Boerner}, G. 1998, \apj, 494, 1

\bibitem[{{Kaiser}(1987)}]{Kaiser1987}
{Kaiser}, N. 1987, \mnras, 227, 1

\bibitem[{{Kessler} {et~al.}(2009){Kessler}, {Becker}, {Cinabro}, {Vanderplas},
  {Frieman}, {Marriner}, {Davis}, {Dilday}, {Holtzman}, {Jha}, {Lampeitl},
  {Sako}, {Smith}, {Zheng}, {Nichol}, {Bassett}, {Bender}, {Depoy}, {Doi},
  {Elson}, {Filippenko}, {Foley}, {Garnavich}, {Hopp}, {Ihara}, {Ketzeback},
  {Kollatschny}, {Konishi}, {Marshall}, {McMillan}, {Miknaitis}, {Morokuma},
  {M{\"o}rtsell}, {Pan}, {Prieto}, {Richmond}, {Riess}, {Romani}, {Schneider},
  {Sollerman}, {Takanashi}, {Tokita}, {van der Heyden}, {Wheeler}, {Yasuda}, \&
  {York}}]{Kessler2009}
{Kessler}, R., {Becker}, A.~C., {Cinabro}, D., {et~al.} 2009, \apjs, 185, 32

\bibitem[{{Kiakotou} {et~al.}(2008){Kiakotou}, {Elgar{\o}y}, \&
  {Lahav}}]{Kiakotou2008}
{Kiakotou}, A., {Elgar{\o}y}, {\O}., \& {Lahav}, O. 2008, Physical Review D,
  77, 063005

\bibitem[{{Komatsu} {et~al.}(2011){Komatsu}, {Smith}, {Dunkley}, {Bennett},
  {Gold}, {Hinshaw}, {Jarosik}, {Larson}, {Nolta}, {Page}, {Spergel},
  {Halpern}, {Hill}, {Kogut}, {Limon}, {Meyer}, {Odegard}, {Tucker}, {Weiland},
  {Wollack}, \& {Wright}}]{Komatsu2011}
{Komatsu}, E., {Smith}, K.~M., {Dunkley}, J., {et~al.} 2011, \apjs, 192, 18

\bibitem[{{Kowalski} {et~al.}(2008){Kowalski}, {Rubin}, {Aldering},
  {Agostinho}, {Amadon}, {Amanullah}, {Balland}, {Barbary}, {Blanc}, {Challis},
  {Conley}, {Connolly}, {Covarrubias}, {Dawson}, {Deustua}, {Ellis}, {Fabbro},
  {Fadeyev}, {Fan}, {Farris}, {Folatelli}, {Frye}, {Garavini}, {Gates},
  {Germany}, {Goldhaber}, {Goldman}, {Goobar}, {Groom}, {Haissinski}, {Hardin},
  {Hook}, {Kent}, {Kim}, {Knop}, {Lidman}, {Linder}, {Mendez}, {Meyers},
  {Miller}, {Moniez}, {Mour{\~a}o}, {Newberg}, {Nobili}, {Nugent}, {Pain},
  {Perdereau}, {Perlmutter}, {Phillips}, {Prasad}, {Quimby}, {Regnault},
  {Rich}, {Rubenstein}, {Ruiz-Lapuente}, {Santos}, {Schaefer}, {Schommer},
  {Smith}, {Soderberg}, {Spadafora}, {Strolger}, {Strovink}, {Suntzeff},
  {Suzuki}, {Thomas}, {Walton}, {Wang}, {Wood-Vasey}, {Yun}, \& {Supernova
  Cosmology Project}}]{Kowalski2008}
{Kowalski}, M., {Rubin}, D., {Aldering}, G., {et~al.} 2008, \apj, 686, 749

\bibitem[{{Leauthaud} {et~al.}(2012){Leauthaud}, {Tinker}, {Bundy}, {Behroozi},
  {Massey}, {Rhodes}, {George}, {Kneib}, {Benson}, {Wechsler}, {Busha},
  {Capak}, {Cort{\^e}s}, {Ilbert}, {Koekemoer}, {Le F{\`e}vre}, {Lilly},
  {McCracken}, {Salvato}, {Schrabback}, {Scoville}, {Smith}, \&
  {Taylor}}]{Leauthaud2012}
{Leauthaud}, A., {Tinker}, J., {Bundy}, K., {et~al.} 2012, \apj, 744, 159

\bibitem[{{Li} {et~al.}(2007){Li}, {Mo}, {van den Bosch}, \& {Lin}}]{Li2007}
{Li}, Y., {Mo}, H.~J., {van den Bosch}, F.~C., \& {Lin}, W.~P. 2007, \mnras,
  379, 689

\bibitem[{{Lin} {et~al.}(2011){Lin}, {Dodelson}, {Seo}, {Soares-Santos},
  {Annis}, {Hao}, {Johnston}, {Kubo}, {Reis}, \& {Simet}}]{Lin2011}
{Lin}, H., {Dodelson}, S., {Seo}, H.-J., {et~al.} 2011, ArXiv e-prints

\bibitem[{{Lueker} {et~al.}(2010){Lueker}, {Reichardt}, {Schaffer}, {Zahn},
  {Ade}, {Aird}, {Benson}, {Bleem}, {Carlstrom}, {Chang}, {Cho}, {Crawford},
  {Crites}, {de Haan}, {Dobbs}, {George}, {Hall}, {Halverson}, {Holder},
  {Holzapfel}, {Hrubes}, {Joy}, {Keisler}, {Knox}, {Lee}, {Leitch}, {McMahon},
  {Mehl}, {Meyer}, {Mohr}, {Montroy}, {Padin}, {Plagge}, {Pryke}, {Ruhl},
  {Shaw}, {Shirokoff}, {Spieler}, {Stalder}, {Staniszewski}, {Stark},
  {Vanderlinde}, {Vieira}, \& {Williamson}}]{Lueker2010}
{Lueker}, M., {Reichardt}, C.~L., {Schaffer}, K.~K., {et~al.} 2010, \apj, 719,
  1045

\bibitem[{{Macci{\`o}} {et~al.}(2008){Macci{\`o}}, {Dutton}, \& {van den
  Bosch}}]{Maccio2008}
{Macci{\`o}}, A.~V., {Dutton}, A.~A., \& {van den Bosch}, F.~C. 2008, \mnras,
  391, 1940

\bibitem[{{Macci{\`o}} {et~al.}(2007){Macci{\`o}}, {Dutton}, {van den Bosch},
  {Moore}, {Potter}, \& {Stadel}}]{Maccio2007}
{Macci{\`o}}, A.~V., {Dutton}, A.~A., {van den Bosch}, F.~C., {et~al.} 2007,
  \mnras, 378, 55

\bibitem[{{Mandelbaum} {et~al.}(2006){Mandelbaum}, {Seljak}, {Kauffmann},
  {Hirata}, \& {Brinkmann}}]{Mandelbaum2006}
{Mandelbaum}, R., {Seljak}, U., {Kauffmann}, G., {Hirata}, C.~M., \&
  {Brinkmann}, J. 2006, \mnras, 368, 715

\bibitem[{{Mantz} {et~al.}(2010){Mantz}, {Allen}, {Rapetti}, \&
  {Ebeling}}]{Mantz2010}
{Mantz}, A., {Allen}, S.~W., {Rapetti}, D., \& {Ebeling}, H. 2010, \mnras, 406,
  1759

\bibitem[{{Massey} {et~al.}(2007){Massey}, {Rhodes}, {Leauthaud}, {Capak},
  {Ellis}, {Koekemoer}, {R{\'e}fr{\'e}gier}, {Scoville}, {Taylor}, {Albert},
  {Berg{\'e}}, {Heymans}, {Johnston}, {Kneib}, {Mellier}, {Mobasher},
  {Semboloni}, {Shopbell}, {Tasca}, \& {Van Waerbeke}}]{Massey2007}
{Massey}, R., {Rhodes}, J., {Leauthaud}, A., {et~al.} 2007, \apjs, 172, 239

\bibitem[{{Mo} {et~al.}(2010){Mo}, {van den Bosch}, \& {White}}]{Mo2010}
{Mo}, H., {van den Bosch}, F.~C., \& {White}, S. 2010, {Galaxy Formation and
  Evolution}

\bibitem[{{More}(2011)}]{More2011b}
{More}, S. 2011, \apj, 741, 19

\bibitem[{{More} {et~al.}(2012){More}, {van den Bosch}, {Cacciato}, {More},
  {Mo}, \& {Yang}}]{More2012clf}
{More}, S., {van den Bosch}, F., {Cacciato}, M., {et~al.} 2012, ArXiv e-prints,
  astro-ph.CO/1204.0786

\bibitem[{{More} {et~al.}(2009{\natexlab{a}}){More}, {van den Bosch}, \&
  {Cacciato}}]{More2009b}
{More}, S., {van den Bosch}, F.~C., \& {Cacciato}, M. 2009{\natexlab{a}},
  \mnras, 392, 917

\bibitem[{{More} {et~al.}(2009{\natexlab{b}}){More}, {van den Bosch},
  {Cacciato}, {Mo}, {Yang}, \& {Li}}]{More2009a}
{More}, S., {van den Bosch}, F.~C., {Cacciato}, M., {et~al.}
  2009{\natexlab{b}}, \mnras, 392, 801

\bibitem[{{More} {et~al.}(2011){More}, {van den Bosch}, {Cacciato}, {Skibba},
  {Mo}, \& {Yang}}]{More2011a}
{More}, S., {van den Bosch}, F.~C., {Cacciato}, M., {et~al.} 2011, \mnras, 410,
  210

\bibitem[{{Navarro} {et~al.}(1997){Navarro}, {Frenk}, \& {White}}]{Navarro1997}
{Navarro}, J.~F., {Frenk}, C.~S., \& {White}, S.~D.~M. 1997, \apj, 490, 493

\bibitem[{{Norberg} {et~al.}(2009){Norberg}, {Baugh}, {Gazta{\~n}aga}, \&
  {Croton}}]{Norberg2009}
{Norberg}, P., {Baugh}, C.~M., {Gazta{\~n}aga}, E., \& {Croton}, D.~J. 2009,
  \mnras, 396, 19

\bibitem[{{Norberg} {et~al.}(2002){Norberg}, {Baugh}, {Hawkins}, {Maddox},
  {Madgwick}, {Lahav}, {Cole}, {Frenk}, {Baldry}, {Bland-Hawthorn}, {Bridges},
  {Cannon}, {Colless}, {Collins}, {Couch}, {Dalton}, {De Propris}, {Driver},
  {Efstathiou}, {Ellis}, {Glazebrook}, {Jackson}, {Lewis}, {Lumsden},
  {Peacock}, {Peterson}, {Sutherland}, \& {Taylor}}]{Norberg2002}
{Norberg}, P., {Baugh}, C.~M., {Hawkins}, E., {et~al.} 2002, \mnras, 332, 827

\bibitem[{{Norberg} {et~al.}(2001){Norberg}, {Baugh}, {Hawkins}, {Maddox},
  {Peacock}, {Cole}, {Frenk}, {Bland-Hawthorn}, {Bridges}, {Cannon}, {Colless},
  {Collins}, {Couch}, {Dalton}, {De Propris}, {Driver}, {Efstathiou}, {Ellis},
  {Glazebrook}, {Jackson}, {Lahav}, {Lewis}, {Lumsden}, {Madgwick}, {Peterson},
  {Sutherland}, \& {Taylor}}]{Norberg2001}
{Norberg}, P., {Baugh}, C.~M., {Hawkins}, E., {et~al.} 2001, \mnras, 328, 64

\bibitem[{{Percival} {et~al.}(2007){Percival}, {Cole}, {Eisenstein}, {Nichol},
  {Peacock}, {Pope}, \& {Szalay}}]{Percival2007}
{Percival}, W.~J., {Cole}, S., {Eisenstein}, D.~J., {et~al.} 2007, \mnras, 381,
  1053

\bibitem[{{Percival} {et~al.}(2010){Percival}, {Reid}, {Eisenstein}, {Bahcall},
  {Budavari}, {Frieman}, {Fukugita}, {Gunn}, {Ivezi{\'c}}, {Knapp}, {Kron},
  {Loveday}, {Lupton}, {McKay}, {Meiksin}, {Nichol}, {Pope}, {Schlegel},
  {Schneider}, {Spergel}, {Stoughton}, {Strauss}, {Szalay}, {Tegmark},
  {Vogeley}, {Weinberg}, {York}, \& {Zehavi}}]{Percival2010}
{Percival}, W.~J., {Reid}, B.~A., {Eisenstein}, D.~J., {et~al.} 2010, \mnras,
  401, 2148

\bibitem[{{Perlmutter} {et~al.}(1999){Perlmutter}, {Aldering}, {Goldhaber},
  {Knop}, {Nugent}, {Castro}, {Deustua}, {Fabbro}, {Goobar}, {Groom}, {Hook},
  {Kim}, {Kim}, {Lee}, {Nunes}, {Pain}, {Pennypacker}, {Quimby}, {Lidman},
  {Ellis}, {Irwin}, {McMahon}, {Ruiz-Lapuente}, {Walton}, {Schaefer}, {Boyle},
  {Filippenko}, {Matheson}, {Fruchter}, {Panagia}, {Newberg}, {Couch}, \& {The
  Supernova Cosmology Project}}]{Perlmutter1999}
{Perlmutter}, S., {Aldering}, G., {Goldhaber}, G., {et~al.} 1999, \apj, 517,
  565

\bibitem[{{Riess} {et~al.}(1998){Riess}, {Filippenko}, {Challis},
  {Clocchiatti}, {Diercks}, {Garnavich}, {Gilliland}, {Hogan}, {Jha},
  {Kirshner}, {Leibundgut}, {Phillips}, {Reiss}, {Schmidt}, {Schommer},
  {Smith}, {Spyromilio}, {Stubbs}, {Suntzeff}, \& {Tonry}}]{Riess1998}
{Riess}, A.~G., {Filippenko}, A.~V., {Challis}, P., {et~al.} 1998, \aj, 116,
  1009

\bibitem[{{Rozo} {et~al.}(2010){Rozo}, {Wechsler}, {Rykoff}, {Annis}, {Becker},
  {Evrard}, {Frieman}, {Hansen}, {Hao}, {Johnston}, {Koester}, {McKay},
  {Sheldon}, \& {Weinberg}}]{Rozo2010}
{Rozo}, E., {Wechsler}, R.~H., {Rykoff}, E.~S., {et~al.} 2010, \apj, 708, 645

\bibitem[{{Schrabback} {et~al.}(2010){Schrabback}, {Hartlap}, {Joachimi},
  {Kilbinger}, {Simon}, {Benabed}, {Brada{\v c}}, {Eifler}, {Erben},
  {Fassnacht}, {High}, {Hilbert}, {Hildebrandt}, {Hoekstra}, {Kuijken},
  {Marshall}, {Mellier}, {Morganson}, {Schneider}, {Semboloni}, {van Waerbeke},
  \& {Velander}}]{Schrabback2010}
{Schrabback}, T., {Hartlap}, J., {Joachimi}, B., {et~al.} 2010, \aap, 516, A63

\bibitem[{{Sehgal} {et~al.}(2011){Sehgal}, {Trac}, {Acquaviva}, {Ade},
  {Aguirre}, {Amiri}, {Appel}, {Barrientos}, {Battistelli}, {Bond}, {Brown},
  {Burger}, {Chervenak}, {Das}, {Devlin}, {Dicker}, {Bertrand Doriese},
  {Dunkley}, {D{\"u}nner}, {Essinger-Hileman}, {Fisher}, {Fowler}, {Hajian},
  {Halpern}, {Hasselfield}, {Hern{\'a}ndez-Monteagudo}, {Hilton}, {Hilton},
  {Hincks}, {Hlozek}, {Holtz}, {Huffenberger}, {Hughes}, {Hughes}, {Infante},
  {Irwin}, {Jones}, {Baptiste Juin}, {Klein}, {Kosowsky}, {Lau}, {Limon},
  {Lin}, {Lupton}, {Marriage}, {Marsden}, {Martocci}, {Mauskopf}, {Menanteau},
  {Moodley}, {Moseley}, {Netterfield}, {Niemack}, {Nolta}, {Page}, {Parker},
  {Partridge}, {Reid}, {Sherwin}, {Sievers}, {Spergel}, {Staggs}, {Swetz},
  {Switzer}, {Thornton}, {Tucker}, {Warne}, {Wollack}, \& {Zhao}}]{Sehgal2011}
{Sehgal}, N., {Trac}, H., {Acquaviva}, V., {et~al.} 2011, \apj, 732, 44

\bibitem[{{Seljak} {et~al.}(2005){Seljak}, {Makarov}, {Mandelbaum}, {Hirata},
  {Padmanabhan}, {McDonald}, {Blanton}, {Tegmark}, {Bahcall}, \&
  {Brinkmann}}]{Seljak2005}
{Seljak}, U., {Makarov}, A., {Mandelbaum}, R., {et~al.} 2005, \prd, 71, 043511

\bibitem[{{Sheldon} {et~al.}(2009){Sheldon}, {Johnston}, {Masjedi}, {McKay},
  {Blanton}, {Scranton}, {Wechsler}, {Koester}, {Hansen}, {Frieman}, \&
  {Annis}}]{Sheldon2009}
{Sheldon}, E.~S., {Johnston}, D.~E., {Masjedi}, M., {et~al.} 2009, \apj, 703,
  2232

\bibitem[{{Smith} {et~al.}(2003){Smith}, {Peacock}, {Jenkins}, {White},
  {Frenk}, {Pearce}, {Thomas}, {Efstathiou}, \& {Couchman}}]{Smith2003}
{Smith}, R.~E., {Peacock}, J.~A., {Jenkins}, A., {et~al.} 2003, \mnras, 341,
  1311

\bibitem[{{Spergel} {et~al.}(2007){Spergel}, {Bean}, {Dor{\'e}}, {Nolta},
  {Bennett}, {Dunkley}, {Hinshaw}, {Jarosik}, {Komatsu}, {Page}, {Peiris},
  {Verde}, {Halpern}, {Hill}, {Kogut}, {Limon}, {Meyer}, {Odegard}, {Tucker},
  {Weiland}, {Wollack}, \& {Wright}}]{Spergel2007}
{Spergel}, D.~N., {Bean}, R., {Dor{\'e}}, O., {et~al.} 2007, \apjs, 170, 377

\bibitem[{{Suzuki} {et~al.}(2012){Suzuki}, {Rubin}, {Lidman}, {Aldering},
  {Amanullah}, {Barbary}, {Barrientos}, {Botyanszki}, {Brodwin}, {Connolly},
  {Dawson}, {Dey}, {Doi}, {Donahue}, {Deustua}, {Eisenhardt}, {Ellingson},
  {Faccioli}, {Fadeyev}, {Fakhouri}, {Fruchter}, {Gilbank}, {Gladders},
  {Goldhaber}, {Gonzalez}, {Goobar}, {Gude}, {Hattori}, {Hoekstra}, {Hsiao},
  {Huang}, {Ihara}, {Jee}, {Johnston}, {Kashikawa}, {Koester}, {Konishi},
  {Kowalski}, {Linder}, {Lubin}, {Melbourne}, {Meyers}, {Morokuma}, {Munshi},
  {Mullis}, {Oda}, {Panagia}, {Perlmutter}, {Postman}, {Pritchard}, {Rhodes},
  {Ripoche}, {Rosati}, {Schlegel}, {Spadafora}, {Stanford}, {Stanishev},
  {Stern}, {Strovink}, {Takanashi}, {Tokita}, {Wagner}, {Wang}, {Yasuda},
  {Yee}, \& {Supernova Cosmology Project}}]{Suzuki2012}
{Suzuki}, N., {Rubin}, D., {Lidman}, C., {et~al.} 2012, \apj, 746, 85

\bibitem[{{Swanson} {et~al.}(2010){Swanson}, {Percival}, \&
  {Lahav}}]{Swanson2010}
{Swanson}, M.~E.~C., {Percival}, W.~J., \& {Lahav}, O. 2010, \mnras, 409, 1100

\bibitem[{{Swanson} {et~al.}(2008){Swanson}, {Tegmark}, {Blanton}, \&
  {Zehavi}}]{Swanson2008}
{Swanson}, M.~E.~C., {Tegmark}, M., {Blanton}, M., \& {Zehavi}, I. 2008,
  \mnras, 385, 1635

\bibitem[{{Tegmark} {et~al.}(2004){Tegmark}, {Blanton}, {Strauss}, {Hoyle},
  {Schlegel}, {Scoccimarro}, {Vogeley}, {Weinberg}, {Zehavi}, {Berlind},
  {Budavari}, {Connolly}, {Eisenstein}, {Finkbeiner}, {Frieman}, {Gunn},
  {Hamilton}, {Hui}, {Jain}, {Johnston}, {Kent}, {Lin}, {Nakajima}, {Nichol},
  {Ostriker}, {Pope}, {Scranton}, {Seljak}, {Sheth}, {Stebbins}, {Szalay},
  {Szapudi}, {Verde}, {Xu}, {Annis}, {Bahcall}, {Brinkmann}, {Burles},
  {Castander}, {Csabai}, {Loveday}, {Doi}, {Fukugita}, {Gott}, {Hennessy},
  {Hogg}, {Ivezi{\'c}}, {Knapp}, {Lamb}, {Lee}, {Lupton}, {McKay}, {Kunszt},
  {Munn}, {O'Connell}, {Peoples}, {Pier}, {Richmond}, {Rockosi}, {Schneider},
  {Stoughton}, {Tucker}, {Vanden Berk}, {Yanny}, \& {York}}]{Tegmark2004}
{Tegmark}, M., {Blanton}, M.~R., {Strauss}, M.~A., {et~al.} 2004, \apj, 606,
  702

\bibitem[{{Tinker} {et~al.}(2008){Tinker}, {Kravtsov}, {Klypin}, {Abazajian},
  {Warren}, {Yepes}, {Gottl{\"o}ber}, \& {Holz}}]{Tinker2008}
{Tinker}, J., {Kravtsov}, A.~V., {Klypin}, A., {et~al.} 2008, \apj, 688, 709

\bibitem[{{Tinker} {et~al.}(2007){Tinker}, {Norberg}, {Weinberg}, \&
  {Warren}}]{Tinker2007}
{Tinker}, J.~L., {Norberg}, P., {Weinberg}, D.~H., \& {Warren}, M.~S. 2007,
  \apj, 659, 877

\bibitem[{{Tinker} {et~al.}(2012){Tinker}, {Sheldon}, {Wechsler}, {Becker},
  {Rozo}, {Zu}, {Weinberg}, {Zehavi}, {Blanton}, {Busha}, \&
  {Koester}}]{Tinker2012}
{Tinker}, J.~L., {Sheldon}, E.~S., {Wechsler}, R.~H., {et~al.} 2012, \apj, 745,
  16

\bibitem[{{Tinker} {et~al.}(2005){Tinker}, {Weinberg}, {Zheng}, \&
  {Zehavi}}]{Tinker2005}
{Tinker}, J.~L., {Weinberg}, D.~H., {Zheng}, Z., \& {Zehavi}, I. 2005, \apj,
  631, 41

\bibitem[{{Vallisneri}(2008)}]{Vallisneri2008}
{Vallisneri}, M. 2008, \prd, 77, 042001

\bibitem[{{van den Bosch} {et~al.}(2003){van den Bosch}, {Mo}, \&
  {Yang}}]{vdb2003}
{van den Bosch}, F.~C., {Mo}, H.~J., \& {Yang}, X. 2003, \mnras, 345, 923

\bibitem[{{van den Bosch} {et~al.}(2007){van den Bosch}, {Yang}, {Mo},
  {Weinmann}, {Macci{\`o}}, {More}, {Cacciato}, {Skibba}, \& {Kang}}]{vdb2007}
{van den Bosch}, F.~C., {Yang}, X., {Mo}, H.~J., {et~al.} 2007, \mnras, 376,
  841

\bibitem[{{Vikhlinin} {et~al.}(2009){Vikhlinin}, {Kravtsov}, {Burenin},
  {Ebeling}, {Forman}, {Hornstrup}, {Jones}, {Murray}, {Nagai}, {Quintana}, \&
  {Voevodkin}}]{Vikhlinin2009}
{Vikhlinin}, A., {Kravtsov}, A.~V., {Burenin}, R.~A., {et~al.} 2009, \apj, 692,
  1060

\bibitem[{{Wang} {et~al.}(2008){Wang}, {Yang}, {Mo}, {van den Bosch},
  {Weinmann}, \& {Chu}}]{Wang2008}
{Wang}, Y., {Yang}, X., {Mo}, H.~J., {et~al.} 2008, \apj, 687, 919

\bibitem[{{Yang} {et~al.}(2003){Yang}, {Mo}, \& {van den Bosch}}]{Yang2003}
{Yang}, X., {Mo}, H.~J., \& {van den Bosch}, F.~C. 2003, \mnras, 339, 1057

\bibitem[{{Yang} {et~al.}(2008){Yang}, {Mo}, \& {van den Bosch}}]{Yang2008a}
{Yang}, X., {Mo}, H.~J., \& {van den Bosch}, F.~C. 2008, \apj, 676, 248

\bibitem[{{Yang} {et~al.}(2009){Yang}, {Mo}, \& {van den Bosch}}]{Yang2009}
{Yang}, X., {Mo}, H.~J., \& {van den Bosch}, F.~C. 2009, \apj, 695, 900

\bibitem[{{Yoo} {et~al.}(2006){Yoo}, {Tinker}, {Weinberg}, {Zheng}, {Katz}, \&
  {Dav{\'e}}}]{Yoo2006}
{Yoo}, J., {Tinker}, J.~L., {Weinberg}, D.~H., {et~al.} 2006, \apj, 652, 26

\bibitem[{{York} {et~al.}(2000){York}, {Adelman}, {Anderson}, {Anderson},
  {Annis}, {Bahcall}, {Bakken}, {Barkhouser}, {Bastian}, {Berman}, {Boroski},
  {Bracker}, {Briegel}, {Briggs}, {Brinkmann}, {Brunner}, {Burles}, {Carey},
  {Carr}, {Castander}, {Chen}, {Colestock}, {Connolly}, {Crocker}, {Csabai},
  {Czarapata}, {Davis}, {Doi}, {Dombeck}, {Eisenstein}, {Ellman}, {Elms},
  {Evans}, {Fan}, {Federwitz}, {Fiscelli}, {Friedman}, {Frieman}, {Fukugita},
  {Gillespie}, {Gunn}, {Gurbani}, {de Haas}, {Haldeman}, {Harris}, {Hayes},
  {Heckman}, {Hennessy}, {Hindsley}, {Holm}, {Holmgren}, {Huang}, {Hull},
  {Husby}, {Ichikawa}, {Ichikawa}, {Ivezi{\'c}}, {Kent}, {Kim}, {Kinney},
  {Klaene}, {Kleinman}, {Kleinman}, {Knapp}, {Korienek}, {Kron}, {Kunszt},
  {Lamb}, {Lee}, {Leger}, {Limmongkol}, {Lindenmeyer}, {Long}, {Loomis},
  {Loveday}, {Lucinio}, {Lupton}, {MacKinnon}, {Mannery}, {Mantsch}, {Margon},
  {McGehee}, {McKay}, {Meiksin}, {Merelli}, {Monet}, {Munn}, {Narayanan},
  {Nash}, {Neilsen}, {Neswold}, {Newberg}, {Nichol}, {Nicinski}, {Nonino},
  {Okada}, {Okamura}, {Ostriker}, {Owen}, {Pauls}, {Peoples}, {Peterson},
  {Petravick}, {Pier}, {Pope}, {Pordes}, {Prosapio}, {Rechenmacher}, {Quinn},
  {Richards}, {Richmond}, {Rivetta}, {Rockosi}, {Ruthmansdorfer}, {Sandford},
  {Schlegel}, {Schneider}, {Sekiguchi}, {Sergey}, {Shimasaku}, {Siegmund},
  {Smee}, {Smith}, {Snedden}, {Stone}, {Stoughton}, {Strauss}, {Stubbs},
  {SubbaRao}, {Szalay}, {Szapudi}, {Szokoly}, {Thakar}, {Tremonti}, {Tucker},
  {Uomoto}, {Vanden Berk}, {Vogeley}, {Waddell}, {Wang}, {Watanabe},
  {Weinberg}, {Yanny}, \& {Yasuda}}]{York2000}
{York}, D.~G., {Adelman}, J., {Anderson}, Jr., J.~E., {et~al.} 2000, \aj, 120,
  1579

\bibitem[{{Zehavi} {et~al.}(2011){Zehavi}, {Zheng}, {Weinberg}, {Blanton},
  {Bahcall}, {Berlind}, {Brinkmann}, {Frieman}, {Gunn}, {Lupton}, {Nichol},
  {Percival}, {Schneider}, {Skibba}, {Strauss}, {Tegmark}, \&
  {York}}]{Zehavi2011}
{Zehavi}, I., {Zheng}, Z., {Weinberg}, D.~H., {et~al.} 2011, \apj, 736, 59

\bibitem[{{Zheng} {et~al.}(2005){Zheng}, {Berlind}, {Weinberg}, {Benson},
  {Baugh}, {Cole}, {Dav{\'e}}, {Frenk}, {Katz}, \& {Lacey}}]{Zheng2005}
{Zheng}, Z., {Berlind}, A.~A., {Weinberg}, D.~H., {et~al.} 2005, \apj, 633, 791

\bibitem[{{Zheng} {et~al.}(2007){Zheng}, {Coil}, \& {Zehavi}}]{Zheng2007}
{Zheng}, Z., {Coil}, A.~L., \& {Zehavi}, I. 2007, \apj, 667, 760

\end{thebibliography}

\appendix
\section{Fisher information matrix}
\label{sec:appendix}

The Fisher information matrix, defined as the (negative of the) second
derivative of the log-likelihood surface, can be used to calculate the
constraints on the parameters of our model, given the observational
data. The Fisher information matrix can be expressed in terms of the
derivatives of the observables with respect to the dimensionless
parameters $\lambda_i$ (see Eqs.~\ref{eq:dimpar} and \ref{eq:fish2}).
For our analysis, we have assumed that the observational constraints
have fixed fractional accuracy. Ignoring the covariance between data
points for simplicity, the dimensionless Fisher information matrix is
given by
\begin{equation}
\tilde{F}_{ij} =  \sum_k \frac{1}{f_{\mu,k}^2}\left( \frac{\partial \ln
\mu_k}{\partial \lambda_i} \right)\left( \frac{\partial \ln \mu_k}{\partial
\lambda_j} \right) \,,
\end{equation}
where $f_{\mu,k}$ is the fractional accuracy of the $k$-th data
constraint and $\left[{\partial \ln \mu_k}/{\partial \lambda}\right]$
is the logarithmic derivatives of the observable with respect to
$\lambda_i$. In this appendix, we present the logarithmic derivatives
of the luminosity function, the galaxy-galaxy clustering signal and
galaxy-galaxy lensing signal with respect to some of our primary model
parameters, which enter the Fisher information matrix defined above.

\begin{figure*}
\centerline{\psfig{figure=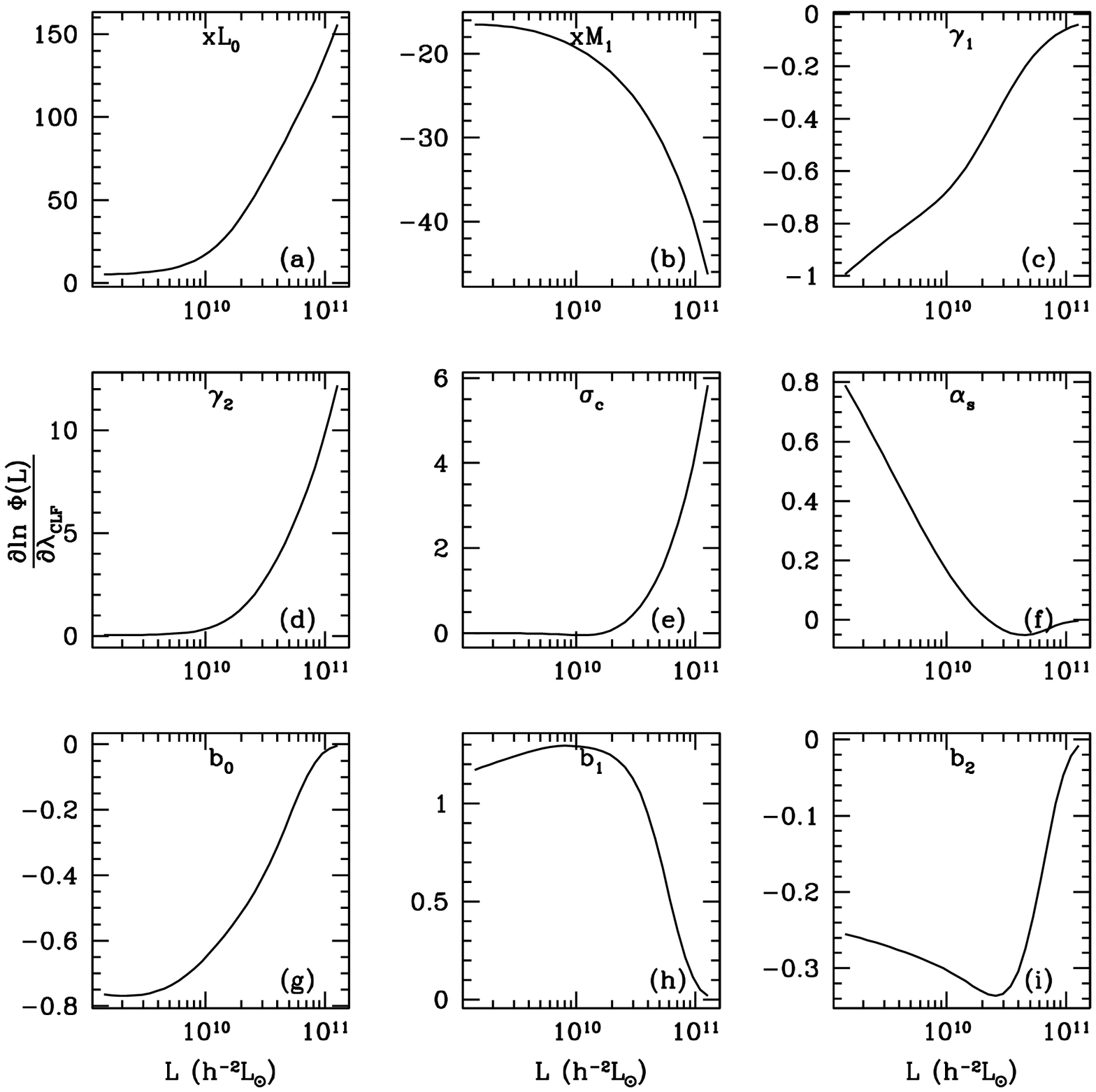,width=0.8\hdsize}}
\caption{ The logarithmic derivatives of the luminosity function with
  respect to the dimensionless CLF parameter $\lambda$. The CLF
  parameter corresponding to each logarithmic derivative is indicated
  at the top of each panel.  }
\label{figLF_CLF}
\end{figure*}
\begin{figure*}
\centerline{\psfig{figure=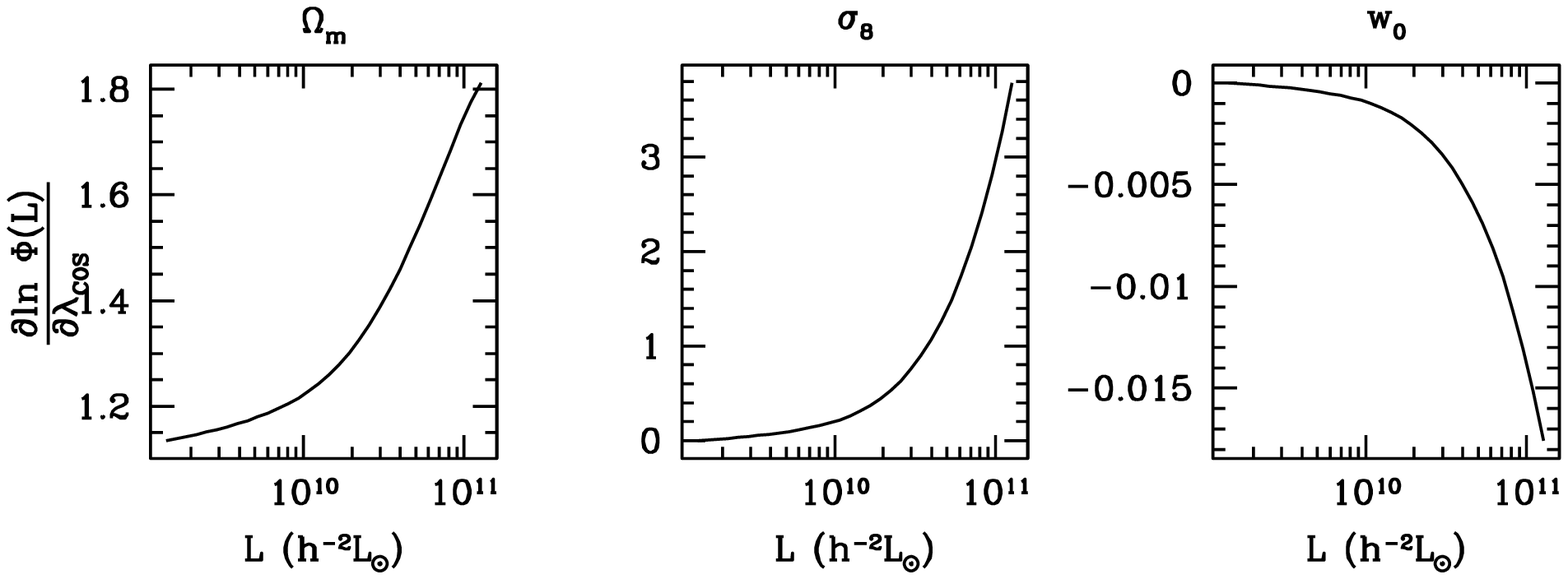,width=0.8\hdsize}}
\caption{ The logarithmic derivatives of the luminosity function with
  respect to the dimensionless cosmological parameters $\lambda$.  The
  cosmological parameter corresponding to each logarithmic derivative
  is indicated at the top of each panel.  }
\label{figLF_Cosmo}
\end{figure*}
\begin{figure*}
\centerline{\psfig{figure=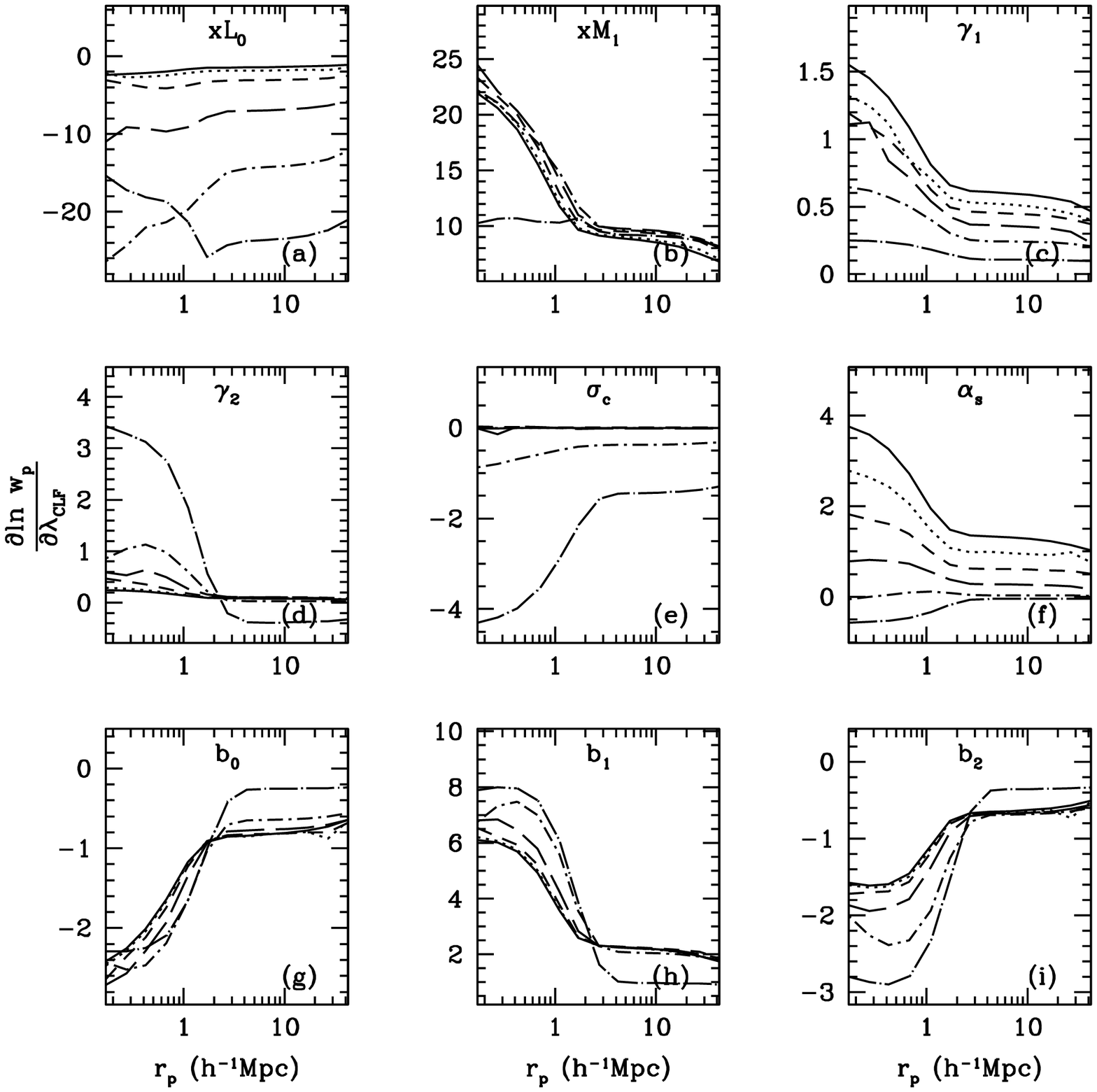,width=0.8\hdsize}}
\caption{ The logarithmic derivatives of the galaxy clustering signal
  with respect to the dimensionless CLF parameter $\lambda$. Different
  line types correspond to the six different luminosity bins. The
  faintest bin is shown using solid line, while the brightest bin is
  shown using a dot-long-dashed line. The CLF parameter corresponding
  to each logarithmic derivative is indicated at the top of each
  panel.  }
\label{figWp_CLF}
\end{figure*}
\begin{figure*}
\centerline{\psfig{figure=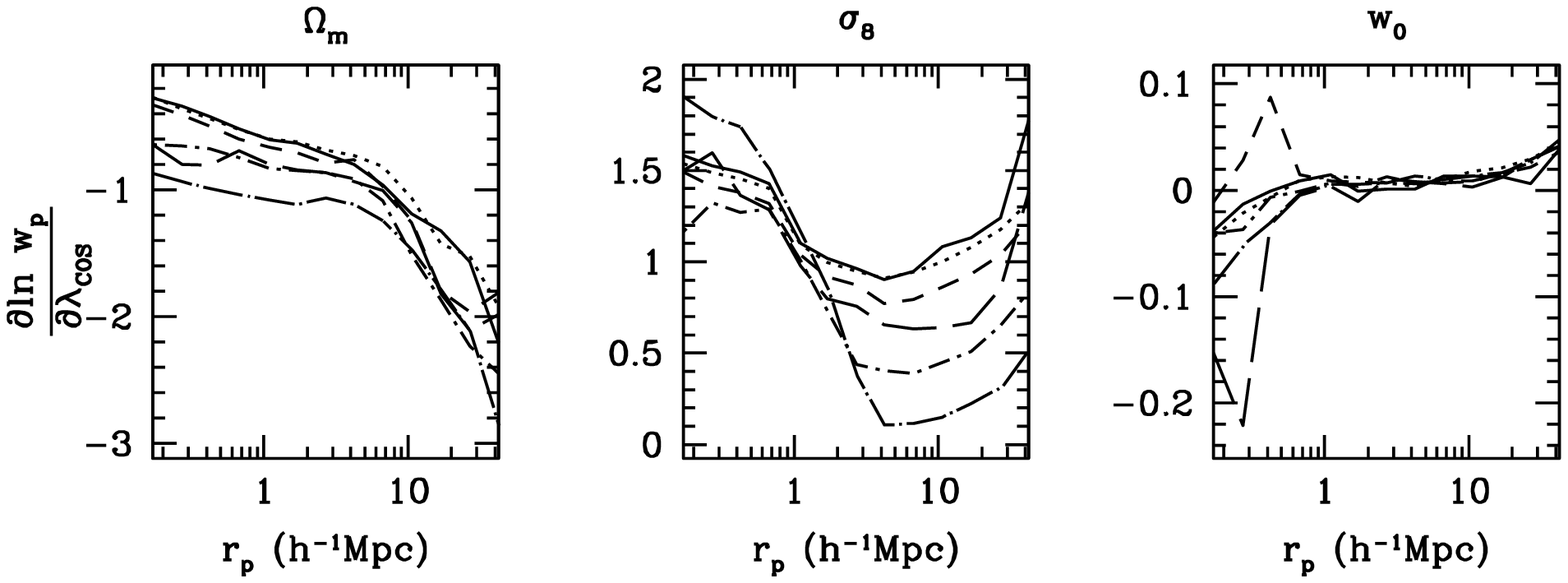,width=0.8\hdsize}}
\caption{ The logarithmic derivatives of the galaxy clustering signal
  with respect to the dimensionless cosmological parameter $\lambda$.
  The cosmological parameter corresponding to each logarithmic
  derivative is indicated at the top of each panel.  The different
  line types correspond to the same bins as that in
  Fig~\ref{figWp_CLF}.  }
\label{figWp_Cosmo}
\end{figure*}
\begin{figure*}
\centerline{\psfig{figure=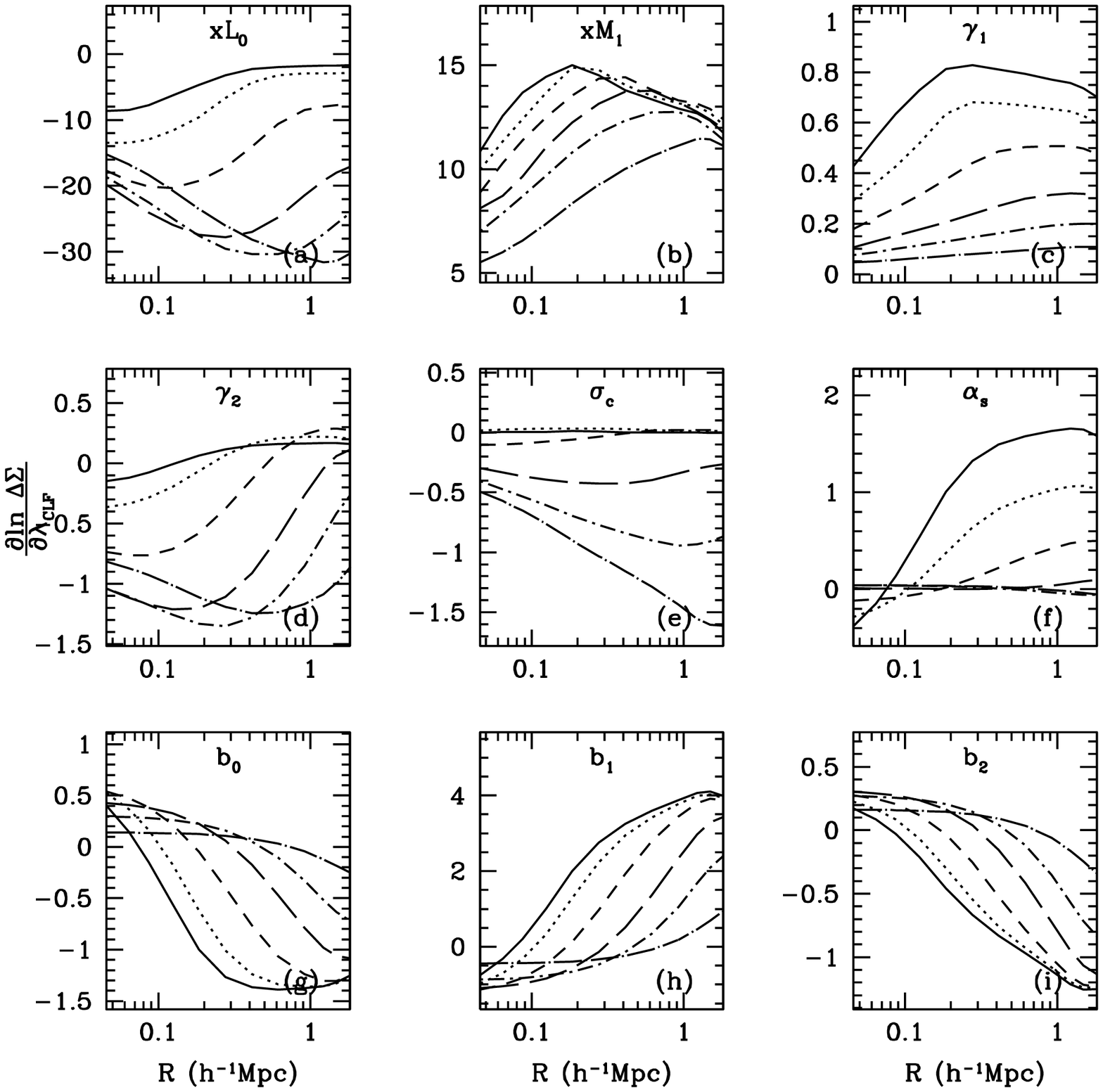,width=0.8\hdsize}}
\caption{ The logarithmic derivatives of the galaxy-galaxy lensing
  signal with respect to the dimensionless CLF parameter $\lambda$.
  The CLF parameter corresponding to each logarithmic derivative is
  indicated at the top of each panel.  The different line types
  correspond to the same bins as that in Fig~\ref{figWp_CLF}.  }
\label{figESD_CLF}
\end{figure*}
\begin{figure*}
\centerline{\psfig{figure=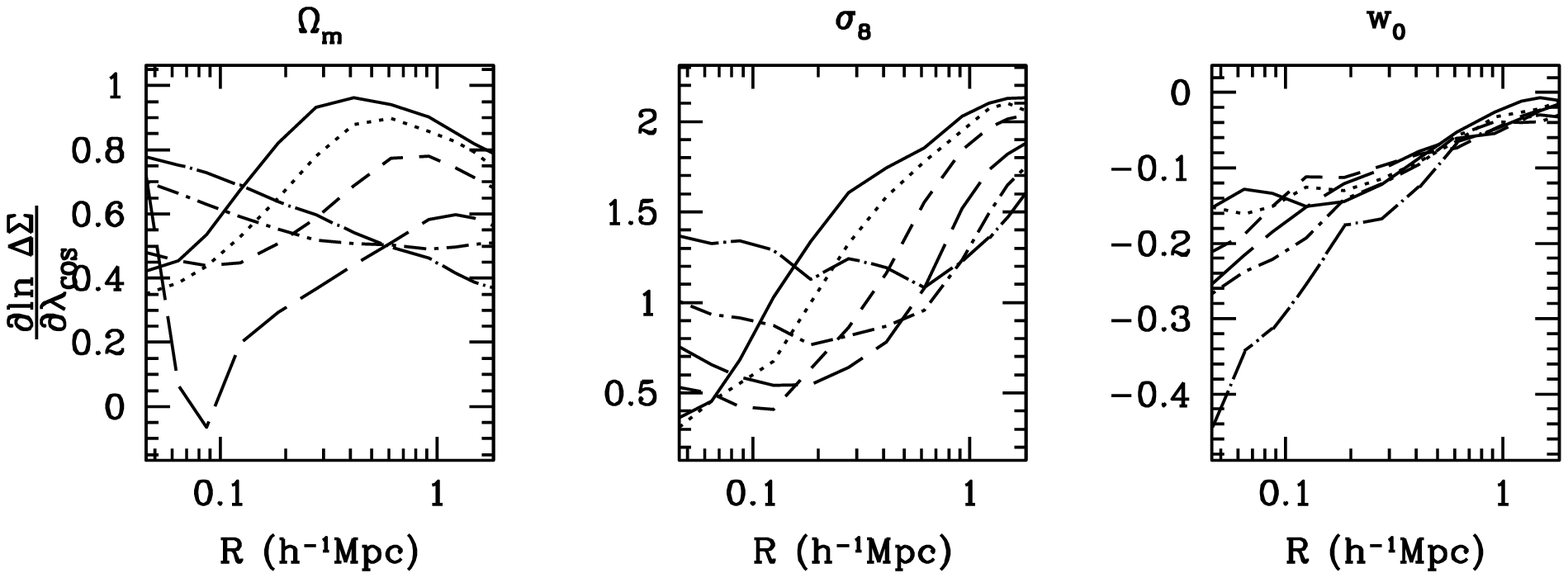,width=0.8\hdsize}}
\caption{ The logarithmic derivatives of the galaxy-galaxy lensing
  signal with respect to the dimensionless cosmological parameter
  $\lambda$.  The cosmological parameter corresponding to each
  logarithmic derivative is indicated at the top of each panel.  The
  different line types correspond to the same bins as that in
  Fig~\ref{figWp_CLF}.  }
\label{figESD_Cosmo}
\end{figure*}

We first focus on the logarithmic derivatives of the luminosity
function with respect to the central and the satellite CLF parameters,
shown in different panels of Fig.~\ref{figLF_CLF}. The luminosity
function of galaxies is always dominated by central galaxies.
Therefore it contains more information about the central CLF
parameters than the satellite CLF parameters, as is evident from a
comparison of the magnitudes of the logarithmic derivatives in panels
(a)-(e) with those in panels (f)-(i). If a positive change in a
parameter causes central galaxies of the same luminosity to live in
lower mass haloes, which are more abundant, it results in an increase
in the galaxy luminosity function and it shows up as a positive
derivative with respect to that parameter. These logarithmic
derivatives with respect to the central CLF parameters are largest at
the bright end (except for the parameter $\gamma_1$ which controls the
low mass end of the halo mass luminosity relation). On the other hand,
the logarithmic derivatives with respect to the satellite CLF
parameters are largest at the faint end because the fractional
contribution of satellite galaxies (satellite fraction) to the
luminosity function increases  as we go to fainter and fainter
galaxies. The luminosity function gives a large amount of information
about the parameters $L_0$ and $M_1$, which determine the pivot points
of the $\tilde{L_\rmc}(M)$ relation. It is interesting to note that
that the derivatives with respect to x$L_0$ and x$M_1$ are opposite in
sign which points to an interesting degeneracy: increasing x$L_0$ can be
compensated by increasing the parameter x$M_1$ if all other parameters
are kept fixed. 

The cosmological parameters affect the luminosity function by changing
the halo mass function. As can be seen from panels (a)-(c) of
Fig~\ref{figLF_Cosmo}, most of the information about the cosmological
parameters comes from the bright end of the galaxy luminosity
function. This is expected because the change in the halo mass
function due to the cosmological parameters is largest at the most
massive end. The derivatives with respect to $\Omega_\rmm$ and $\sigma_8$
have the same sign and are only slightly different in shape. Hence a
positive change in $\Omega_\rmm$ can be countered with a negative change
in $\sigma_8$ and vice versa. The weakest constraints are on the
parameter $w_0$. The parameter $w_0$ mainly affects the luminosity function
by changing the growth of structure and thus the power spectrum. We
have assumed that the luminosity function is only measured at a single
redshift $z=0.1$ from SDSS data, thus the leverage in redshift to
propagate changes to the growth of structure are not very large. This
results in very weak sensitivity to $w_0$.

The logarithmic derivatives of the projected galaxy clustering with
respect to the CLF parameters are shown in Fig.~\ref{figWp_CLF}. On large
scales, the clustering of galaxies mainly arises due to the two halo
term which is sensitive to the clustering of haloes in which these
galaxies reside. Since more massive haloes are more strongly clustered,
a change in a parameter which causes galaxies to reside on average in
more massive haloes will result in an increase in the clustering of
galaxies on large scales. On small scales, the one halo terms 
(central-satellite and the satellite-satellite) dominate the
clustering of galaxies. The satellite distribution in more
massive haloes is less concentrated but the number of satellites in
these haloes is larger and there are competing effects that can result
in changes in the clustering of galaxies in directions opposite to the
trend on large scales (see e.g. panel [d]). The trends in the
logarithmic derivatives, as shown in Fig.~\ref{figWp_CLF}, on average show
that the small scale clustering of galaxies has the most information
about the halo occupation parameters. The galaxies in the brightest
bins show a large derivative with respect to the central CLF
parameters (barring $\gamma_1$ as expected). The satellite CLF
parameters that determine the normalization of the satellite CLF are
almost equally sensitive to all luminosity bins. The logarithmic
derivatives are largest for the parameters x$L_0$ and x$M_1$. Although
the derivatives are opposite in sign, the shapes of these derivatives
as a function of scale are quite different from each other, thus
providing an avenue to constrain the parameters separately.

Comparisons of the logarithmic derivatives of the galaxy clustering
signal with respect to the cosmological parameters, shown in
Fig.~\ref{figWp_Cosmo}, to the derivatives shown in
Fig.~\ref{figLF_Cosmo}, give an interesting insight. Unlike the
logarithmic derivatives of the luminosity function, the derivatives
with respect to $\Omega_\rmm$ and $\sigma_8$ have opposite signs,
which suggests that a combination of the two observables adds a lot of
information about these parameters.  The derivatives of the galaxy
clustering signal with respect to the parameter $\Omega_\rmm$ increase
on large scales. The parameter $\sigma_8$ controls the concentration
of dark matter haloes and therefore the distribution of the satellite
galaxies. Therefore it affects both the small and the large scales.
Galaxy clustering on large scales is also sensitive to the parameter
$w_0$, because it affects the growth of structure, while on small
scales the sensitivity is expected due to the changes to the
concentration-mass relation.

The information contained in the galaxy-galaxy lensing signal, as
shown by the logarithmic derivatives in Fig.~\ref{figESD_CLF} and
Fig.~\ref{figESD_Cosmo}, is dominated by the one halo terms as the
signal has only been measured out to a couple of megaparsecs. The
derivatives with respect to the CLF parameters show quite different
dependencies on scale compared to the clustering data, with a large
amout of information coming from the scales of the order of a
megaparsec. Most of the trends are fairly similar to the trends in the
logarithmic derivatives of the galaxy-galaxy clustering data and
depend upon how the CLF parameters change the halo occupation
distribution of galaxies. The logarithmic derivatives with respect to
the cosmological parameters show that measuring the galaxy-galaxy
signal to larger scales can certainly add more information. We have
quantified how the constraints on the cosmological parameters improve
by using the galaxy-galaxy lensing measured to $\sim30\,\mpch$ signal
in Section~\ref{sec:constraints}. 

\section{Confidence intervals from the Fisher matrix}
The inverse of the Fisher information matrix gives the covariance with
which the parameters of the model can be constrained given the data.
Here, we briefly present the procedure to obtain the confidence
intervals in a two dimensional parameter space given this covariance
matrix. Let $P(x,y)$ denote the posterior probability distribution for
two parameters of interest $x$ and $y$ after marginalizing over the
rest of the parameter set.  For simplicity, let us assume that the
constraints on these two parameters are uncorrelated, which implies
that the subset of the full covariance matrix corresponding to these
two parameters is a diagonal matrix given by
\begin{equation}
C =
\begin{pmatrix}
\sigma^2_{xx} & 0 \\
0 & \sigma^2_{yy}
\end{pmatrix}.
\end{equation}
In case the covariance matrix is not diagonal, one can always
diagonalise it by rotating the co-ordinate system and aligning the
axes with the eigenvectors of this covariance matrix. Let us assume
that the posterior distribution $P(x,y)$ is Gaussian and without loss
of generality, let us assume that it is centred at $(0,0)$, such that
\begin{equation}
    P(x,y) =
    \frac{1}{2\pi\,[\sigma_{xx}\sigma_{yy}]}
\exp\left(-\left[\frac{x^2}{2\sigma_{xx}^2} +
\frac{y^2}{2\sigma_{yy}^2}\right]\right) \,.
\end{equation}
Changing variables to the dimensionless quantities $\xi'$ and $\theta$
defined such that
\begin{equation}
x=\sigma_{xx}\xi'\cos(\theta); \,\,y=\sigma_{yy}\xi'\sin(\theta)\,,
\end{equation}
gives
\begin{eqnarray}
P(x,y)\,\drm x\,\drm y &=& P(\xi',\theta) \drm \xi'\,\drm \theta \,,\\ &=&
\frac{\xi'\,\drm\xi'\,\drm\theta}{2\pi}\,\exp\left(\frac{-\xi'^2}{2}\right)\,.
\end{eqnarray}
The probability that the true model lies within a region enclosed by
the iso-probability contour $\xi$ is then given by
\begin{equation}
P_\xi=\int_{0}^{2\pi} \int_{0}^{\xi}
P(\xi',\theta)\,\drm\xi'\,\drm\theta=1-\exp\left[\frac{-\xi^2}{2}\right]\,.
\end{equation}
The iso-probability contours $\xi=1.5096$, $2.4477$ and $3.0349$
correspond to $P_\xi=0.68$, $0.95$ and $0.99$, respectively. These
values of $\xi$ correspond to 68, 95 and 99 percent confidence
ellipses in the $x-y$ plane with principal axes of length equal to
$\xi\sigma_{xx}$ and $\xi\sigma_{yy}$. 

To summarize, the step-by-step procedure we use to obtain the
confidence ellipses is as follows: (i) invert the entire $n_{\rm
par}\times n_{\rm par}$ Fisher matrix to obtain the covariance matrix,
(ii) choose the $2\times 2$ subset of the covariance matrix
corresponding to the two parameters of interest, (iii) find the
eigenvalues and eigenvectors of this submatrix, (iv) draw 68, 95 and
99 percent confidence ellipses with the axes aligned to the
eigenvectors, and with axes lengths equal to $1.5096$, $2.4477$ and
$3.0349$ times the square root of the corresponding eigenvalues found
in step (iii), respectively. Note that, this procedure can be
trivially generalized to obtain confidence ellipsoids in a subspace
spanned by any number of parameters.

\end{document}